\documentclass[aip,jmp,amsmath,amssymb,reprint]{revtex4-2}

\usepackage{graphicx} 
\usepackage{dcolumn}  
\usepackage{bm}       
\usepackage{mathrsfs}
\usepackage{tensor}
\usepackage{hyperref}
\usepackage{tikz}     
\usepackage{color}
\usetikzlibrary{arrows.meta,decorations.markings,calc}
\hypersetup{
    colorlinks=true,
    linkcolor=blue,
    filecolor=magenta,      
    urlcolor=blue,
    citecolor=blue,
}
\usetikzlibrary{arrows.meta,decorations.markings}
\begin{document}

\preprint{AIP/123-QED}

\title{Analytical derivation of long-term dephasing caused by phase transitions in the context of Kerr black holes}

\author{Jingxu Wu}
\email{wuxj@my.msu.ru}
\thanks{Corresponding author.}
\affiliation{Faculty of Physics, Lomonosov Moscow State University, Leninskie Gory, Moscow 119991, Russia}

\author{Liangyu Luo}
\email{niu985@sina.com}
\thanks{Co-corresponding author.}
\affiliation{International School, I.M. Sechenov First Moscow State Medical University, Moscow 119048, Russia}

\author{Jie Shi}
\affiliation{Faculty of Physics, Lomonosov Moscow State University, Leninskie Gory, Moscow 119991, Russia}

\date{\today}

\begin{abstract}
Extreme Mass Ratio Inspirals (EMRIs) constitute a prime target for future space-based gravitational-wave observatories such as LISA. In this paper, we analytically investigate the long-term phase shift (dephasing) in the gravitational wave signal induced by a first-order quantum chromodynamics (QCD) phase transition within a neutron star orbiting a supermassive Kerr black hole. By modeling the transition from a hadronic phase to a quark core phase, we quantify the sudden change in the tidal deformability ($\Lambda$) of the secondary object. Utilizing the Teukolsky formalism and Post-Newtonian expansions, we derive a strict analytical scaling law for the accumulated dephasing. We demonstrate that the Kerr spin parameter $a$ and the critical phase transition orbital velocity $v_c$ significantly amplify the dephasing effect. Our analytical framework provides a robust tool for probing the non-perturbative QCD equation of state at high baryon densities using gravitational wave astronomy.
\end{abstract}

\maketitle

\section{\label{sec:intro}Introduction}

Extreme-mass-ratio inspirals (EMRIs), in which a compact stellar remnant spirals into a massive black hole, are among the most theoretically rich sources in gravitational-wave astrophysics. Their importance is rooted in a combination of features that is essentially unique: they probe the deepest strong-field region of a Kerr spacetime; they remain in band for long durations; and they accumulate an enormous orbital phase, rendering them exquisitely sensitive to even weak departures from a purely point-particle inspiral. For this reason EMRIs have long been recognized as one of the flagship science targets of the Laser Interferometer Space Antenna (LISA), whose low-frequency band is precisely matched to the slow, relativistic inspiral of compact objects into massive black holes \cite{LISAESA,Hughes2000,Hughes2001,BarackPound2019,Khalvati2024,SearchEMRI2025}.

This very strength is also the central challenge. Because EMRIs accumulate such a large number of gravitational-wave cycles, the observable signal depends not only on the background spacetime geometry, but also on any persistent mechanism capable of inducing a secular phase drift. In the vacuum problem, this is the domain of black-hole perturbation theory, self-force methods, and increasingly accurate relativistic waveform models. In realistic astrophysical settings, however, a number of additional channels may perturb the phasing: environmental forces, departures from general relativity, finite-size effects of the secondary, and internal-structure effects if the smaller body is a neutron star rather than a black hole \cite{BarackPound2019,Khalvati2024,SearchEMRI2025}. The present paper focuses on the last of these possibilities and asks a specific question: can a phase transition in the dense interior of a neutron-star secondary leave a coherent dephasing imprint on an EMRI waveform in Kerr spacetime?

The underlying physical motivation is compelling. Quantum chromodynamics predicts that at sufficiently high baryon density hadronic matter may undergo a transition to deconfined quark matter. Yet the phase structure of cold, strongly interacting matter at neutron-star densities remains one of the least understood sectors of modern physics. Terrestrial heavy-ion experiments probe high temperature and relatively small baryon chemical potential, while the interiors of neutron stars access the opposite regime: low temperature, extreme density, and strong beta equilibrium. In this sense compact stars provide a natural laboratory for nonperturbative QCD under conditions inaccessible on Earth. The question is not simply whether quark matter exists somewhere inside neutron stars, but whether any such transition can be inferred observationally from macroscopic observables such as masses, radii, tidal deformabilities, or oscillation spectra \cite{HanSteiner2019,Takatsy2023,RaithelMost2023PRL,RaithelMost2023PRD,Roy2024,Li2025QCDPT}.

Over the last several years it has become increasingly clear that tidal observables are among the most promising quantities for addressing this question. A strong first-order transition can significantly alter the mass--radius relation and reduce the stellar tidal deformability, while smoother hybrid-star constructions may still remain compatible with present multimessenger constraints. At the same time, recent studies have emphasized that tidal measurements alone may not uniquely determine the existence or nature of a phase transition, because qualitatively different equations of state can in some cases mimic similar inspiral-scale tidal signatures. This ambiguity is not a weakness of tidal physics itself, but rather a reminder that one must carefully understand how microscopic structure is transmitted into the waveform phase \cite{HanSteiner2019,RaithelMost2023PRL,RaithelMost2023PRD,Takatsy2023,Roy2024}. A closely related recent development is the proposal that interface-mode resonances in inspiralling neutron stars could provide a more direct smoking-gun signature of a first-order transition. That mechanism is dynamical and resonant in character. The mechanism explored in the present work is different: here we isolate the adiabatic, secular dephasing channel associated with a transition-induced change in the effective tidal response \cite{Counsell2025,Pereira2025}.

The central conceptual step of this paper is to exploit the natural separation of scales in the EMRI problem. The orbital motion is governed by the background Kerr spacetime and is treated as an adiabatic sequence of circular equatorial geodesics. The dense-matter microphysics enters only through the tidal deformability of the neutron-star secondary, which is allowed to change when the inspiral crosses a critical regime. In this way the nonperturbative QCD transition is not modeled as a modification of the spacetime geometry itself, but as a perturbative source term in the dissipative evolution. This makes it possible to keep the strong-field Kerr structure exact at the background level while still propagating a microphysical transition into a waveform-level phase observable. Technically, the resulting framework is especially well suited to symbolic manipulation and post-Newtonian re-expansion, which allows one to extract closed analytical scaling laws for the accumulated phase shift.

Our specific scenario is the following. A neutron star of mass $\mu$ inspirals around a supermassive Kerr black hole of mass $M$ with mass ratio $q=\mu/M\ll1$. As the orbit shrinks, the tidal field acting on the neutron star strengthens. If the stellar core crosses the threshold for deconfinement, the internal composition changes from a hadronic phase to a quark-core phase, or more generally passes through a finite-width mixed phase. This transition modifies the stellar compactness and Love number, and hence changes the dimensionless tidal deformability $\Lambda$. Once inserted into the radiative flux, this modification generates a long-term difference between the phase history of the transition-bearing inspiral and that of a purely hadronic baseline. Because EMRIs accumulate phase over an exceptionally long time, even a perturbatively small change in the tidal sector can in principle integrate into a measurable dephasing.

The main purpose of the present paper is to derive this dephasing analytically and to expose its dependence on the two physically most important control parameters: the orbital location of the transition and the spin of the central Kerr black hole. To keep the analysis transparent, we work in the circular equatorial limit and adopt an adiabatic treatment of the inspiral. Within that setting we construct the Kerr phase kernel from the exact circular-orbit energy and a Teukolsky-inspired flux expansion, and then inject the phase transition through an effective tidal source term. We consider both an idealized sharp first-order transition and a finite-width mixed-phase interpolation, derive the associated dephasing formulas, and use them to generate parameter-space scans and waveform-level diagnostics.

The results obtained here support three main conclusions. First, the phase-transition contribution can be written in a factorized form that cleanly separates Kerr strong-field dynamics from dense-matter microphysics. Second, the accumulated phase shift is highly sensitive to the orbital location of the transition and is significantly modulated by the black-hole spin through both the phase kernel and the strong-field endpoint of the inspiral. Third, the resulting signal appears primarily as a secular waveform dephasing rather than as an amplitude anomaly, making EMRIs a particularly natural arena in which to search for this effect. In this sense the problem studied here sits at the intersection of two frontiers: strong-field general relativity and the phase structure of dense QCD matter.

The paper is organized as follows. In Sec.~\ref{sec:dynamics} we review the circular equatorial Kerr inspiral dynamics and introduce the point-particle flux kernel used throughout the analysis. In Sec.~\ref{sec:qcd_model} we formulate the effective microphysical model of the hadron--quark transition in terms of the tidal deformability, considering both sharp and mixed-phase profiles. In Sec.~\ref{sec:derivation} we derive the analytical dephasing formulas and obtain their post-Newtonian and spin-expanded scaling laws. In Sec.~\ref{sec:numerical} we present numerical realizations of the analytical model, including parameter scans and time-domain waveform comparisons. We summarize the physical implications and outline future extensions in Sec.~\ref{sec:conclusion}.
\section{\label{sec:dynamics}EMRI Dynamics in Kerr Spacetime}

Extreme-mass-ratio inspirals are naturally described within a two-timescale framework: on the short timescale the secondary follows a geodesic of the background Kerr geometry, while on the long radiation-reaction timescale the orbital constants drift under gravitational-wave dissipation \cite{Hughes2000,Hughes2001,BarackPound2019}. In the present work we restrict attention to equatorial circular motion, which is the cleanest setting for isolating the dephasing generated by an internal phase transition of the neutron-star secondary. This restriction is not merely pedagogical. Circular--equatorial Kerr inspirals remain a standard strong-field benchmark for black-hole perturbation theory, relativistic waveform generation, and recent Lorenz-gauge metric-reconstruction developments \cite{Teukolsky1973,Dolan2024}. Throughout this section we use geometrized units $G=c=1$.

\subsection{Background Orbit and Radiative Fluxes}

Consider a compact object of mass $\mu$ orbiting a Kerr black hole of mass $M$ and dimensionless spin $\chi=a/M$, with mass ratio
\begin{equation}
q \equiv \frac{\mu}{M}\ll 1.
\end{equation}
For equatorial circular geodesics it is convenient to absorb the orbital orientation into the signed spin parameter
\begin{equation}
\hat a \equiv \sigma \chi, 
\qquad
\sigma=
\begin{cases}
+1, & \text{prograde},\\
-1, & \text{retrograde}.
\end{cases}
\end{equation}
The orbital angular frequency measured at infinity is then
\begin{equation}
\Omega
=
\frac{1}{M\left[(r/M)^{3/2}+\hat a\right]},
\label{eq:Omega_r_section2}
\end{equation}
which is the standard Bardeen--Press--Teukolsky result for Kerr circular motion \cite{BardeenPressTeukolsky1972}. For later post-Newtonian manipulations we introduce the invariant frequency parameter
\begin{equation}
v \equiv (M\Omega)^{1/3},
\label{eq:v_def_section2}
\end{equation}
together with
\begin{equation}
u(v,\hat a)=\frac{v^2}{\left(1-\hat a v^3\right)^{2/3}}=\frac{M}{r}.
\label{eq:u_of_v_section2}
\end{equation}
In terms of $u$, the specific orbital energy $\tilde E\equiv E/\mu$ takes the exact form
\begin{equation}
\tilde E(v,\hat a)
=
\frac{1-2u+\hat a\,u^{3/2}}
{\sqrt{1-3u+2\hat a\,u^{3/2}}}.
\label{eq:Etilde_exact_section2}
\end{equation}
This compact representation is particularly useful because it makes the Kerr geodesic structure explicit while keeping the later PN re-expansion algebraically tractable.

In the adiabatic regime the inspiral is driven by the balance between the slow secular loss of orbital energy and the gravitational-wave flux radiated to infinity and into the event horizon,
\begin{equation}
\frac{dE}{dt}=-\mathcal{F}_{\rm PP},
\qquad
\mathcal{F}_{\rm PP}=\mathcal{F}^{\infty}+\mathcal{F}^{H},
\label{eq:balance_pp_section2}
\end{equation}
where the subscript ``PP'' indicates the point-particle baseline before tidal corrections are included \cite{Hughes2000,Hughes2001}. In black-hole perturbation theory these fluxes are computed from solutions of the Teukolsky equation for the spin-weighted Weyl scalar $\psi_4$ \cite{Teukolsky1973}. For the present circular--equatorial problem it is sufficient to regard $\mathcal{F}_{\rm PP}$ as the sum of asymptotic fluxes extracted from the Teukolsky amplitudes at null infinity and on the horizon. Recent fully relativistic waveform constructions for circular equatorial EMRIs, as well as new Kerr metric-perturbation calculations in Lorenz gauge, reinforce the usefulness of this setup as a controlled strong-field baseline \cite{Dolan2024}.

For analytic estimates we normalize the flux by $\mu^2$,
\begin{equation}
\hat{\mathcal F}_{\rm PP}\equiv \frac{\mathcal F_{\rm PP}}{\mu^2},
\end{equation}
and adopt a PN-consistent Kerr expansion of the form
\begin{align}
\hat{\mathcal F}_{\rm PP}(v,\hat a)
&=
\frac{32}{5}v^{10}
\Bigg[
1
-\frac{1247}{336}v^2
+\left(4\pi-\frac{11}{4}\hat a\right)v^3
\nonumber\\
&\qquad
+\left(-\frac{44711}{9072}+\frac{33}{16}\hat a^2\right)v^4
+\left(-\frac{8191\pi}{672}-\frac{59}{16}\hat a\right)v^5
+O(v^6)
\Bigg],
\label{eq:Fpp_section2}
\end{align}
which captures the leading Schwarzschild terms together with the lowest spin-dependent Kerr corrections relevant to our analytic dephasing model \cite{Fujita2015,Blanchet2024PN}. The explicit coefficients beyond the orders displayed are not needed in the main text; when higher-order terms are required, they are supplied by the symbolic pipeline described later.

Combining Eqs.~\eqref{eq:Etilde_exact_section2} and \eqref{eq:Fpp_section2}, we define the point-particle phase kernel
\begin{equation}
K(v,\hat a)
\equiv
-\frac{v^3}{q}\,
\frac{d\tilde E/dv}{\hat{\mathcal F}_{\rm PP}(v,\hat a)}.
\label{eq:K_def_section2}
\end{equation}
This quantity isolates the purely relativistic Kerr background and will serve as the backbone of the dephasing calculation. In particular, once the tidal sector is written multiplicatively as a perturbation of the flux, the phase-transition contribution can be expressed compactly as the convolution of $K(v,\hat a)$ with the tidal source. This separation between a ``vacuum Kerr kernel'' and a ``microphysical tidal insertion'' is one of the main structural simplifications of the present approach.

\subsection{Tidal Coupling and Teukolsky Formalism}

The neutron-star secondary responds to the external tidal field by developing an induced quadrupole moment,
\begin{equation}
Q_{ij}=-\lambda\,\mathcal E_{ij},
\label{eq:Qij_section2}
\end{equation}
where $\lambda$ is the dimensional tidal deformability and $\mathcal E_{ij}$ is the electric part of the Weyl tensor evaluated in the local rest frame of the star \cite{FengLyuYang2021,FengLyuYang2022}. It is convenient to trade $\lambda$ for the dimensionless tidal deformability
\begin{equation}
\Lambda
\equiv
\frac{\lambda}{\mu^5}
=
\frac{2}{3}k_2\,C^{-5},
\qquad
C\equiv \frac{\mu}{R},
\label{eq:Lambda_def_section2}
\end{equation}
with $k_2$ the quadrupolar Love number and $R$ the stellar radius. In the present problem the key physical assumption is that the hadron-to-quark transition modifies $(R,k_2)$, and hence $\Lambda$, while the orbital motion remains accurately described as an adiabatic inspiral in the fixed Kerr background. At the level of accuracy pursued here, the phase transition is therefore introduced as a controlled perturbation of the flux sector rather than as a deformation of the background spacetime itself.

Within black-hole perturbation theory the radiative degrees of freedom are governed by the Teukolsky master equation,
\begin{equation}
\mathcal{T}_s[\Psi_s]=4\pi \Sigma\,\mathcal{S}_s,
\label{eq:teuk_master_section2}
\end{equation}
with spin weight $s=-2$ for the outgoing gravitational sector \cite{Teukolsky1973}. The source term $\mathcal{S}_s$ depends on the stress-energy content of the secondary and therefore inherits the tidal multipole structure generated by Eq.~\eqref{eq:Qij_section2}. In a strict first-order treatment the tidal correction can be organized as a small multiplicative deformation of the point-particle flux,
\begin{equation}
\hat{\mathcal F}_{\rm total}(v,\hat a,\Lambda)
=
\hat{\mathcal F}_{\rm PP}(v,\hat a)
\left[
1+\Lambda\,\mathcal H_{\rm tidal}(v,\hat a)
\right],
\label{eq:Ftotal_section2}
\end{equation}
where $\mathcal H_{\rm tidal}$ is dimensionless. For equatorial circular inspirals this structure is directly motivated by the BHPT+PN analyses of tidal couplings in Kerr spacetime, which show that the leading tidal phase correction is frequency dependent, agrees with PN theory in the weak field, and remains only mildly spin dependent over the relevant parameter range \cite{FengLyuYang2021,FengLyuYang2022}.

In keeping with the expected quadrupolar counting, we parameterize the tidal sector as
\begin{equation}
\mathcal H_{\rm tidal}(v,\hat a)
=
\kappa_0 v^{10}
\left(
1+h_2 v^2+h_{3s}\hat a\,v^3+h_4 v^4+\cdots
\right),
\label{eq:Htidal_section2}
\end{equation}
where $\kappa_0$ fixes the leading tidal strength and the coefficients $(h_2,h_{3s},h_4)$ encode higher-order nonspinning and spin-tidal corrections. Equation~\eqref{eq:Htidal_section2} is deliberately written in a model-independent way: it preserves the correct 5PN tidal scaling while allowing the explicit coefficients to be supplied either by analytic BHPT calculations or by calibrated numerical Teukolsky fluxes. This is precisely the level of generality needed for the present paper. The dephasing formula derived in Sec.~\ref{sec:derivation} depends only on the structure of Eqs.~\eqref{eq:K_def_section2} and \eqref{eq:Htidal_section2}, not on the details of how the coefficients are obtained.

Finally, it is useful to emphasize the hierarchy of approximations adopted in this section. The background motion is treated as a sequence of circular Kerr geodesics; dissipation is captured through Teukolsky fluxes; the phase transition enters only through the induced change in the stellar tidal deformability; and all such tidal effects are kept perturbative relative to the point-particle inspiral. This hierarchy is well aligned with the modern EMRI program, in which the geodesic-plus-flux picture provides the leading-order adiabatic scaffold, while progressively refined self-force, metric-perturbation, and waveform models are added systematically on top of it \cite{BarackPound2019,Dolan2024}. In the next section we specify the microphysical model for $\Lambda(v)$ and show how a first-order transition or a finite-width mixed phase is injected into the dephasing integral.

\section{\label{sec:qcd_model}Modeling the QCD Phase Transition}

The central physical assumption of the present work is that the dense-matter transition inside the neutron-star secondary does not need to be resolved microscopically in real time in order to leave a measurable imprint on the gravitational-wave phase. For the purpose of inspiral phasing, it is sufficient to identify the macroscopic response quantity through which the internal QCD dynamics enters the orbital evolution. In the adiabatic regime considered here, that quantity is the tidal deformability. The role of the hadron--quark transition is therefore encoded in a controlled modification of the stellar tidal response, while the orbital motion itself continues to be described by circular Kerr inspiral dynamics as developed in Sec.~\ref{sec:dynamics}.

This effective perspective is well motivated by the present status of dense-matter phenomenology. Current astrophysical analyses admit hybrid-star equations of state compatible with modern mass, radius, and tidal-deformability constraints, but they do not yet uniquely determine whether the relevant high-density transition is a strong first-order jump, a finite-width mixed phase, or a smoother crossover-like rearrangement \cite{HanSteiner2019,Takatsy2023,RaithelMost2023PRL,Roy2024}. From the viewpoint of gravitational-wave phasing, what matters most at leading order is not the full microscopic free-energy functional, but rather how the phase transition modifies the stellar compactness and Love number, and hence the dimensionless tidal deformability.

\begin{figure*}[htbp]
\centering
\begin{tikzpicture}[
    scale=0.88,
    transform shape,
    line cap=round,
    line join=round,
    >=Latex,
    every node/.style={font=\small}
]

\definecolor{hadronblue}{RGB}{70,130,180}
\definecolor{quarkred}{RGB}{220,20,60}
\definecolor{kerrblack}{RGB}{25,25,25}
\definecolor{ergogray}{RGB}{210,210,210}
\definecolor{softgray}{RGB}{120,120,120}
\definecolor{accentorange}{RGB}{235,145,40}
\definecolor{deepviolet}{RGB}{110,80,170}

\tikzset{
    flowarrow/.style={
        postaction={decorate},
        decoration={
            markings,
            mark=at position 0.83 with {\arrow{Latex[length=3mm,width=2mm]}}
        }
    },
    lab/.style={
        fill=white,
        rounded corners=2pt,
        inner sep=2pt,
        opacity=0.96,
        text opacity=1
    }
}

\shade[inner color=gray!8, outer color=ergogray, opacity=0.85]
    (0,0) ellipse (2.00 and 1.28);

\draw[gray!70, densely dotted, line width=1.0pt] (0,0) circle (1.52);
\node[lab, anchor=west] at (1.60,-0.98) {$r_{\rm ISCO}(\hat a)$};

\draw[quarkred!80!black, dash pattern=on 5pt off 3pt, line width=1.2pt]
    (0,0) circle (2.90);
\node[lab, text=quarkred!85!black, anchor=west] at (3.00,-1.60) {$r_c$};

\fill[kerrblack] (0,0) circle (0.78);
\node[text=white, font=\small\bfseries] at (0,0) {$M,\hat a$};

\node[lab, text=gray!60!black, anchor=west] at (-4.60,1.45) {ergoregion};
\draw[gray!60!black, -Latex, line width=0.9pt]
    (-3.45,1.28) -- (-1.55,0.92);

\pgfmathsetmacro{\tcrit}{9.90}

\draw[hadronblue, line width=1.7pt, flowarrow, variable=\x,
domain=0:\tcrit, samples=220, smooth]
plot ({(5.4 - 0.25*\x)*cos(deg(\x))},
      {0.60*(5.4 - 0.25*\x)*sin(deg(\x))});

\draw[quarkred, line width=1.9pt, variable=\xx,
domain=\tcrit:14.0, samples=200, smooth]
plot ({(5.4 - 0.25*\xx)*cos(deg(\xx))},
      {0.60*(5.4 - 0.25*\xx)*sin(deg(\xx))});
\pgfmathsetmacro{\xc}{(5.4 - 0.25*\tcrit)*cos(deg(\tcrit))}
\pgfmathsetmacro{\yc}{0.60*(5.4 - 0.25*\tcrit)*sin(deg(\tcrit))}
\fill[accentorange] (\xc,\yc) circle (2.0pt);
\draw[accentorange, line width=0.8pt] (\xc,\yc) circle (3.6pt);

\node[lab, text=accentorange!90!black, anchor=west] at (-4.85,-2.45)
    {transition onset};
\draw[accentorange!90!black, -Latex, line width=1pt]
    (-3.20,-2.28) -- (\xc-0.08,\yc-0.08);

\node[lab, text=hadronblue!90!black, anchor=west] at (3.45,2.35)
    {\textbf{hadronic response}};
\draw[hadronblue!80!black, -Latex, line width=1pt]
    (3.35,2.08) -- (2.10,1.42);

\node[lab, text=quarkred!90!black, anchor=west] at (2.75,0.62)
    {\textbf{quark-core response}};
\draw[quarkred!80!black, -Latex, line width=1pt]
    (2.65,0.38) -- (1.35,0.12);

\node[lab, anchor=west] at (-4.85,2.82)
    {\textbf{EMRI inspiral trajectory}};
\draw[softgray, -Latex, line width=1pt]
    (-2.95,2.62) -- (-1.08,1.88);

\begin{scope}[shift={(-6.4,1.55)}]
    \fill[hadronblue!18] (0,0) circle (0.84);
    \draw[hadronblue!85!black, line width=1.2pt] (0,0) circle (0.84);

    \draw[hadronblue!70!black, line width=1.0pt]
        plot[smooth cycle, tension=0.9]
        coordinates {(0,0.84) (0.55,0.45) (0.90,0) (0.55,-0.45) (0,-0.84) (-0.64,-0.48) (-0.78,0) (-0.64,0.48)};

    \node[font=\footnotesize\bfseries, text=hadronblue!90!black] at (0,-1.24) {hadronic NS};

    \node[lab, anchor=west] at (1.10,0.34) {$\Lambda_{\rm H}$};
    \draw[hadronblue!80!black, -Latex, line width=0.9pt] (0.98,0.24) -- (0.60,0.16);

    \draw[hadronblue!70!black, dashed, line width=0.9pt] (0.92,-0.04) -- (2.75,-0.34);
\end{scope}

\begin{scope}[shift={(6.4,1.55)}]
    \fill[quarkred!14] (0,0) circle (0.84);
    \draw[quarkred!85!black, line width=1.2pt] (0,0) circle (0.84);

    \draw[quarkred!75!black, line width=1.0pt]
        plot[smooth cycle, tension=0.95]
        coordinates {(0,0.82) (0.50,0.42) (0.80,0) (0.50,-0.42) (0,-0.82) (-0.54,-0.42) (-0.68,0) (-0.54,0.42)};

    \fill[deepviolet!80] (0,0) circle (0.30);
    \draw[deepviolet!95!black, line width=0.8pt] (0,0) circle (0.30);

    \node[font=\footnotesize\bfseries, text=quarkred!90!black] at (0,-1.24) {hybrid / quark-core NS};

    \node[lab, anchor=east] at (-1.08,0.54) {$\Lambda_{\rm Q}$};
    \draw[quarkred!80!black, -Latex, line width=0.9pt] (-0.96,0.42) -- (-0.64,0.30);

    \node[lab, anchor=east] at (-1.08,-0.18) {$\Lambda_{\rm Q}<\Lambda_{\rm H}$};
    \draw[quarkred!80!black, -Latex, line width=0.9pt] (-0.96,-0.05) -- (-0.40,-0.05);

    \draw[quarkred!70!black, dashed, line width=0.9pt] (-0.92,-0.10) -- (-2.78,-0.72);
\end{scope}

\node[lab, align=center] at (0,-3.95)
{
effective tidal jump at $r=r_c$\\
$\Delta\Lambda \equiv \Lambda_{\rm Q}-\Lambda_{\rm H}<0$
};

\draw[accentorange!85!black, line width=1.0pt, -Latex] (-0.50,-3.50) -- (\xc-0.05,-0.48);
\draw[accentorange!85!black, line width=1.0pt, -Latex] (0.50,-3.50) -- (\xc+0.05,-0.32);

\end{tikzpicture}
\caption{\label{fig:schematic}
Schematic illustration of an EMRI in which the neutron-star secondary undergoes a microphysical transition during inspiral around a Kerr black hole. Outside the critical radius $r_c$, the companion remains in a hadronic configuration with tidal deformability $\Lambda_{\rm H}$ (blue segment). Once the orbit enters the critical regime, the star develops a denser quark core and the effective tidal response changes to $\Lambda_{\rm Q}<\Lambda_{\rm H}$ (red segment). The strong-field endpoint is set by $r_{\rm ISCO}(\hat a)$, while the phase evolution is correspondingly modified by the transition-induced jump $\Delta\Lambda<0$.}
\end{figure*}
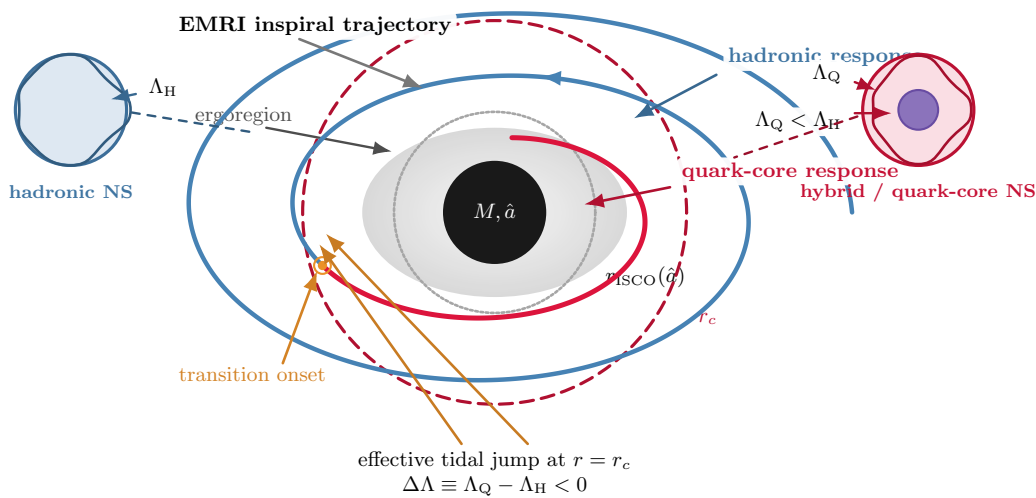

\subsection{Tidal deformability as the effective microphysical variable}

We characterize the stellar response through the dimensionless quadrupolar tidal deformability
\begin{equation}
\Lambda
=
\frac{\lambda}{\mu^5}
=
\frac{2}{3}k_2 C^{-5},
\qquad
C\equiv \frac{\mu}{R},
\label{eq:Lambda_qcdmodel}
\end{equation}
where $\lambda$ is the dimensional tidal deformability, $k_2$ is the quadrupolar Love number, $R$ is the stellar radius, and $C$ is the compactness. In this parametrization all information about the high-density equation of state enters the inspiral problem through the pair $(R,k_2)$ evaluated along the relevant stellar branch.

A hadron--quark transition generally makes the inner core more compressible and the equilibrium configuration more compact. At fixed stellar mass, this tends to reduce the radius and suppress the tidal response. We therefore write
\begin{equation}
\Lambda_H \rightarrow \Lambda_Q,
\qquad
\Delta\Lambda \equiv \Lambda_Q-\Lambda_H,
\label{eq:DeltaLambda_def}
\end{equation}
with
\begin{equation}
\Delta\Lambda < 0
\end{equation}
in the canonical scenario of interest. This sign convention will be important later, because it determines the sign of the accumulated dephasing: once the star becomes less deformable, the tidal contribution to the dissipation weakens relative to the purely hadronic baseline.

The advantage of formulating the microphysics in terms of $\Lambda$ is twofold. First, it provides a direct bridge to gravitational-wave observables, since tidal effects enter the phase evolution already at the inspiral level. Second, it isolates the phase-transition signal from the unresolved details of the microscopic conversion mechanism. In other words, different hybrid-star constructions may lead to distinct internal compositions while still producing similar effective values of $(\Lambda_H,\Lambda_Q)$ at the level relevant for phasing. This is precisely why an explicitly parameterized framework is needed before one attempts any model-dependent inference \cite{RaithelMost2023PRL,RaithelMost2023PRD}.

\subsection{Sharp first-order transition}

The simplest idealization is a sharp first-order transition, corresponding to a Maxwell-type construction in which the star remains on the hadronic branch until a critical internal threshold is reached, after which the equilibrium response jumps abruptly to a quark-core branch. In the inspiral problem, the most economical description is to treat this threshold as being crossed at a critical orbital velocity $v_c$, or equivalently at a critical orbital frequency $\Omega_c$. We therefore model the response as
\begin{equation}
\Lambda(v)
=
\Lambda_H+\Delta\Lambda\,\Theta(v-v_c),
\label{eq:Lambda_step_qcdmodel}
\end{equation}
where $\Theta$ is the Heaviside step function.

Equation~\eqref{eq:Lambda_step_qcdmodel} should be interpreted as an effective orbital representation of a microphysical threshold crossing. The underlying trigger may be formulated in several equivalent ways: a critical central pressure, a critical baryon chemical potential, a threshold core density, or a sufficiently strong tide-induced restructuring of the stellar interior. Since our observable is the long-term gravitational-wave phase rather than the local conversion dynamics, the detailed trigger mechanism need not be specified at the level of the main formalism. What matters is that, on the radiation-reaction timescale, the change in the tidal response is sufficiently abrupt that the step-function approximation captures its net effect on the inspiral.

For later use it is convenient to define the deviation from the purely hadronic baseline,
\begin{equation}
\delta\Lambda(v)\equiv \Lambda(v)-\Lambda_H.
\label{eq:deltalambda_general}
\end{equation}
In the sharp-transition model this becomes
\begin{equation}
\delta\Lambda_{\rm step}(v)=\Delta\Lambda\,\Theta(v-v_c).
\label{eq:deltalambda_step}
\end{equation}
This representation will enter the phase integral in Sec.~\ref{sec:derivation}, where the lower limit of the phase-transition contribution is automatically shifted from the detector entrance frequency to the transition point $v_c$.

It is worth emphasizing that a sharp transition is not merely a mathematical convenience. Strong first-order phase transitions can induce pronounced changes in the mass--radius relation and in the tidal deformability, and may even generate disconnected stable branches or ``twin-star''-like behavior in suitable regions of parameter space \cite{HanSteiner2019,Takatsy2023}. From the standpoint of waveform phasing, such a scenario is precisely the one most likely to yield a clear differential signal relative to a purely hadronic inspiral.

\subsection{Mixed-phase and finite-width transition}

A discontinuous jump is, however, not the only viable possibility. If the interface tension is small, if the matter satisfies global rather than local charge-neutrality conditions over part of the transition layer, or if the conversion process is effectively smeared over a finite density interval, then the transition may proceed through a mixed phase rather than an infinitesimally thin boundary. Even if the underlying microscopic equation of state contains a first-order transition, the mapping from internal thermodynamic variables to the orbital parameter $v$ may smooth the effective inspiral response.

To model this situation while retaining analytic control, we introduce a finite transition window bounded by $v_h$ and $v_q$, corresponding respectively to the onset and completion of the mixed phase. Defining
\begin{equation}
y \equiv \frac{v-v_h}{v_q-v_h},
\qquad
S(y)=3y^2-2y^3,
\label{eq:smoothstep_qcdmodel}
\end{equation}
we write
\begin{equation}
\Lambda(v)=
\begin{cases}
\Lambda_H, & v\le v_h, \\[4pt]
\Lambda_H+(\Lambda_Q-\Lambda_H)\,S(y), & v_h < v < v_q, \\[6pt]
\Lambda_Q, & v\ge v_q.
\end{cases}
\label{eq:Lambda_mixed_qcdmodel}
\end{equation}
The corresponding deviation from the hadronic baseline is
\begin{equation}
\delta\Lambda_{\rm mix}(v)=
\begin{cases}
0, & v\le v_h, \\[4pt]
(\Lambda_Q-\Lambda_H)\,S(y), & v_h < v < v_q, \\[6pt]
\Lambda_Q-\Lambda_H, & v\ge v_q.
\end{cases}
\label{eq:deltalambda_mixed}
\end{equation}

This form has several advantages. First, both $\Lambda(v)$ and $d\Lambda/dv$ are continuous at the endpoints, which avoids spurious cusps in the dephasing kernel. Second, the interpolation remains polynomial inside the transition interval, so the symbolic phase integrals can still be carried out analytically order by order in the PN expansion. Third, the width
\begin{equation}
\Delta v \equiv v_q-v_h
\end{equation}
provides a direct phenomenological measure of how extended the microphysical conversion is in orbital space. In the limit $\Delta v\to 0$, Eq.~\eqref{eq:Lambda_mixed_qcdmodel} reduces smoothly to the sharp-transition model of Eq.~\eqref{eq:Lambda_step_qcdmodel}.

The astrophysical importance of this distinction is substantial. Recent Bayesian and waveform-oriented studies have shown that smooth hybrid equations of state can remain consistent with present data while producing inspiral signatures that are less dramatic than those generated by a strong discontinuity \cite{Roy2024,Li2025QCDPT}. Conversely, low-density transitions may mimic purely hadronic tidal behavior closely enough to create degeneracies in the inverse problem, even when the underlying microphysics is qualitatively different \cite{RaithelMost2023PRL,RaithelMost2023PRD}. This is precisely why it is useful to treat the mixed-phase width as an independent effective parameter rather than assuming an abrupt conversion from the outset.

\subsection{From microphysics to the dephasing source term}

The net effect of the QCD phase transition is therefore summarized by the map
\begin{equation}
\{\Lambda_H,\Lambda_Q,v_c\}
\quad \text{or} \quad
\{\Lambda_H,\Lambda_Q,v_h,v_q\}
\longrightarrow
\delta\Lambda(v),
\label{eq:micro_map}
\end{equation}
which is then inserted into the tidal sector of the phase evolution. This separation between background Kerr dynamics and microphysical response is the conceptual core of the paper: the orbital motion supplies the universal relativistic kernel, while the dense-matter physics enters only through the state-dependent function $\delta\Lambda(v)$.

It is also important to clarify what this paper does \emph{not} attempt to capture. We do not include dynamical mode excitations associated with a sharp density interface, nor do we model the conversion front as a nonlinear hydrodynamic instability. Those effects may well provide additional observables, and in fact recent work suggests that interface modes could furnish a more direct probe of a genuine first-order phase transition in inspiralling neutron stars \cite{Counsell2025}. Our focus here is narrower but cleaner: we isolate the adiabatic dephasing channel alone and derive its analytic scaling with the transition parameters.

Within this framework, the key observable question is no longer whether quark matter exists in some abstract sense, but whether a change in the effective tidal deformability occurs early enough and strongly enough during the inspiral to accumulate a measurable phase offset. That question is answered by combining the present section with the Kerr phase kernel of Sec.~\ref{sec:dynamics}. In the next section we show explicitly how $\delta\Lambda(v)$ enters the energy-balance equation and derive closed analytic expressions for the phase shift in both the sharp-transition and mixed-phase cases.

\section{\label{sec:derivation}Analytical Derivation of the Dephasing}

Having specified in Sec.~\ref{sec:dynamics} the Kerr background dynamics and in Sec.~\ref{sec:qcd_model} the effective microphysical input through $\delta\Lambda(v)$, we now derive the central observable of this paper: the cumulative gravitational-wave phase shift induced by a QCD phase transition inside the neutron-star secondary. The logic of the derivation is simple but physically important. The inspiral is treated as an adiabatic sequence of circular equatorial Kerr orbits; the baseline radiation reaction is encoded in the point-particle Teukolsky flux; and the phase transition enters only through a perturbative change in the tidal response of the secondary. In this formulation the strong-field geometry and the dense-matter microphysics are cleanly disentangled: the former determines the universal Kerr phase kernel, while the latter appears as a source term that modulates the dissipation.

This separation makes it possible to formulate the dephasing problem at fixed orbital state rather than at fixed coordinate time. For two histories with different radiation-reaction rates, a comparison at the same coordinate time mixes a genuine phase effect with the trivial fact that the two systems have already reached different orbital configurations. The natural gauge-invariant comparison is therefore performed at the same instantaneous orbital frequency, or equivalently at the same PN variable $v=(M\Omega)^{1/3}$. In the adiabatic regime this is precisely the variable in terms of which the phase accumulation can be written most transparently \cite{Hughes2000,Hughes2001,BarackPound2019,MathewsPound2025}.

\subsection{Energy Balance Equation with Phase Transition Source}

We begin with the adiabatic energy-balance law,
\begin{equation}
\frac{dE}{dt}=-\mathcal F_{\rm total},
\label{eq:balance_total_section4}
\end{equation}
where $E=\mu \tilde E$ is the orbital energy and $\mathcal F_{\rm total}$ is the total gravitational-wave luminosity, including both the point-particle baseline and the tidal correction induced by the neutron-star response. Since the orbital phase obeys $d\Phi/dt=\Omega$, one may write
\begin{equation}
\frac{d\Phi}{d\Omega}
=
\frac{\Omega}{\dot\Omega}
=
\Omega\,
\frac{dE/d\Omega}{dE/dt}
=
-\Omega\,
\frac{dE/d\Omega}{\mathcal F_{\rm total}}.
\label{eq:dphidomega_section4}
\end{equation}
Changing variables from $\Omega$ to $v=(M\Omega)^{1/3}$ gives
\begin{equation}
\Omega=\frac{v^3}{M},
\qquad
d\Omega = \frac{3v^2}{M}\,dv,
\end{equation}
and hence
\begin{equation}
\frac{d\Phi}{dv}
=
-\frac{v^3}{q}\,
\frac{d\tilde E/dv}{\hat{\mathcal F}_{\rm total}(v,\hat a,\Lambda)},
\label{eq:dphidv_exact_section4}
\end{equation}
where
\begin{equation}
\hat{\mathcal F}_{\rm total}\equiv \frac{\mathcal F_{\rm total}}{\mu^2},
\qquad
q\equiv \frac{\mu}{M}.
\end{equation}
Equation~\eqref{eq:dphidv_exact_section4} is exact within the adiabatic circular-orbit approximation and makes explicit the hierarchy of scales: the background Kerr geometry enters through $\tilde E(v,\hat a)$, while dissipation appears through the normalized flux.

We now insert the effective tidal decomposition introduced in Sec.~\ref{sec:dynamics},
\begin{equation}
\hat{\mathcal F}_{\rm total}(v,\hat a,\Lambda)
=
\hat{\mathcal F}_{\rm PP}(v,\hat a)
\left[
1+\Lambda(v)\,\mathcal H_{\rm tidal}(v,\hat a)
\right].
\label{eq:Ftotal_factorized_section4}
\end{equation}
It is useful to isolate the purely Kerr background part by defining
\begin{equation}
K(v,\hat a)
\equiv
-\frac{v^3}{q}\,
\frac{d\tilde E/dv}{\hat{\mathcal F}_{\rm PP}(v,\hat a)}.
\label{eq:K_redef_section4}
\end{equation}
Then Eq.~\eqref{eq:dphidv_exact_section4} becomes
\begin{equation}
\frac{d\Phi}{dv}
=
\frac{K(v,\hat a)}
{1+\Lambda(v)\,\mathcal H_{\rm tidal}(v,\hat a)}.
\label{eq:dphidv_factorized_section4}
\end{equation}
This form is central to the whole analysis. It shows explicitly that, to the accuracy pursued here, the entire influence of dense QCD matter enters as a deformation of the denominator of the phase kernel.

Since the tidal sector is perturbative, we work in the regime
\begin{equation}
\big|\Lambda(v)\,\mathcal H_{\rm tidal}(v,\hat a)\big|\ll 1,
\label{eq:small_tidal_parameter_section4}
\end{equation}
and expand
\begin{equation}
\frac{1}{1+\Lambda\mathcal H_{\rm tidal}}
=
1-\Lambda\mathcal H_{\rm tidal}
+O\!\left[(\Lambda\mathcal H_{\rm tidal})^2\right].
\end{equation}
Equation~\eqref{eq:dphidv_factorized_section4} then yields
\begin{equation}
\frac{d\Phi}{dv}
\approx
K(v,\hat a)
\left[
1-\Lambda(v)\,\mathcal H_{\rm tidal}(v,\hat a)
\right].
\label{eq:dphidv_linearized_section4}
\end{equation}
We are not interested in the total phase itself, but in the \emph{difference} between two inspiral histories: one in which the neutron star remains on the hadronic branch throughout the inspiral, and one in which it undergoes the transition described in Sec.~\ref{sec:qcd_model}. Let
\begin{equation}
\delta\Lambda(v)\equiv \Lambda(v)-\Lambda_H
\end{equation}
be the transition-induced departure from the purely hadronic baseline. Subtracting the two phase histories at fixed $v$, one obtains the dephasing source equation
\begin{equation}
\Delta\Phi_{\rm PT}
=
-\int_{v_{\rm in}}^{v_{\rm ISCO}}
K(v,\hat a)\,
\delta\Lambda(v)\,
\mathcal H_{\rm tidal}(v,\hat a)\,dv.
\label{eq:master_dephasing_section4}
\end{equation}
Equation~\eqref{eq:master_dephasing_section4} is the master formula of the paper. It states that the long-term dephasing is obtained by folding the microphysical response function $\delta\Lambda(v)$ against a universal Kerr background kernel.

The practical advantage of Eq.~\eqref{eq:master_dephasing_section4} is that different microphysical scenarios can now be treated by simply replacing $\delta\Lambda(v)$. For the sharp first-order model of Sec.~\ref{sec:qcd_model},
\begin{equation}
\delta\Lambda_{\rm step}(v)
=
\Delta\Lambda\,\Theta(v-v_c),
\label{eq:deltalambda_step_section4}
\end{equation}
and therefore
\begin{equation}
\Delta\Phi_{\rm step}
=
-\Delta\Lambda
\int_{v_c}^{v_{\rm ISCO}}
K(v,\hat a)\,
\mathcal H_{\rm tidal}(v,\hat a)\,dv.
\label{eq:DeltaPhi_step_general_section4}
\end{equation}
For the mixed-phase model,
\begin{equation}
\delta\Lambda_{\rm mix}(v)=
\begin{cases}
0, & v\le v_h,\\[4pt]
(\Lambda_Q-\Lambda_H)\,S\!\left(\dfrac{v-v_h}{v_q-v_h}\right), & v_h<v<v_q,\\[8pt]
\Lambda_Q-\Lambda_H, & v\ge v_q,
\end{cases}
\label{eq:deltalambda_mix_section4}
\end{equation}
with
\begin{equation}
S(y)=3y^2-2y^3,
\qquad
y=\frac{v-v_h}{v_q-v_h},
\end{equation}
one finds
\begin{align}
\Delta\Phi_{\rm mix}
&=
-(\Lambda_Q-\Lambda_H)
\int_{v_h}^{v_q}
K(v,\hat a)\,
S\!\left(\frac{v-v_h}{v_q-v_h}\right)
\mathcal H_{\rm tidal}(v,\hat a)\,dv
\nonumber\\
&\quad
-(\Lambda_Q-\Lambda_H)
\int_{v_q}^{v_{\rm ISCO}}
K(v,\hat a)\,
\mathcal H_{\rm tidal}(v,\hat a)\,dv.
\label{eq:DeltaPhi_mix_general_section4}
\end{align}
The decomposition into a finite-width transition piece and a post-transition tail will later prove useful for understanding how the transition width competes with the strong-field amplification near the end of the inspiral.

\subsection{Analytical Scaling Laws and Spin Dependence}

To obtain closed analytic expressions, we now expand the Kerr orbital quantities and the flux kernel in the small-spin and PN regimes. Writing the signed spin as $\hat a$ and expanding to linear order in $\hat a$, one finds for the specific energy
\begin{align}
\tilde E(v,\hat a)
&=
1
-\frac{1}{2}v^2
+\frac{3}{8}v^4
+\frac{27}{16}v^6
+\frac{675}{128}v^8
+\frac{3969}{256}v^{10}
\nonumber\\
&\quad
-\frac{4}{3}\hat a\,v^5
-4\hat a\,v^7
-\frac{27}{2}\hat a\,v^9
+O(v^{11},\hat a^2),
\label{eq:E_pn_section4}
\end{align}
and therefore
\begin{align}
\frac{d\tilde E}{dv}
&=
-v
+\frac{3}{2}v^3
+\frac{81}{8}v^5
+\frac{675}{16}v^7
+\frac{19845}{128}v^9
\nonumber\\
&\quad
-\frac{20}{3}\hat a\,v^4
-28\hat a\,v^6
-\frac{243}{2}\hat a\,v^8
+O(v^{10},\hat a^2).
\label{eq:dE_pn_section4}
\end{align}
The point-particle flux is expanded in the form
\begin{align}
\hat{\mathcal F}_{\rm PP}(v,\hat a)
&=
\frac{32}{5}v^{10}
\Bigg[
1
-\frac{1247}{336}v^2
+\left(4\pi-\frac{11}{4}\hat a\right)v^3
\nonumber\\
&\qquad
+\left(-\frac{44711}{9072}+\frac{33}{16}\hat a^2\right)v^4
+\left(-\frac{8191\pi}{672}-\frac{59}{16}\hat a\right)v^5
+O(v^6)
\Bigg],
\label{eq:Fpp_repeated_section4}
\end{align}
while the tidal sector is parameterized as
\begin{equation}
\mathcal H_{\rm tidal}(v,\hat a)
=
\kappa_0 v^{10}
\left(
1+h_2 v^2+h_{3s}\hat a\,v^3+h_4 v^4+\cdots
\right).
\label{eq:Htidal_repeated_section4}
\end{equation}
Substituting Eqs.~\eqref{eq:dE_pn_section4} and \eqref{eq:Fpp_repeated_section4} into the definition of the phase kernel, Eq.~\eqref{eq:K_redef_section4}, we obtain
\begin{align}
K(v,\hat a)
&=
\frac{5}{32q}v^{-6}
+\frac{3715}{10752q}v^{-4}
+\left(
-\frac{5\pi}{8q}
+\frac{565}{384q}\hat a
\right)v^{-3}
\nonumber\\
&\quad
+\frac{15293365}{32514048q}v^{-2}
+\left(
-\frac{38645\pi}{21504q}
+\frac{732985}{64512q}\hat a
\right)v^{-1}
+O(v^{0},\hat a^2).
\label{eq:K_pn_section4}
\end{align}
Several features are already visible at this stage. First, the leading kernel scales as $v^{-6}$, which is the familiar balance-law enhancement associated with the long accumulation time of the inspiral. Second, the spin enters not only through the orbital energy but also through the flux, and therefore appears already at the level of the effective dephasing kernel. Third, because the tidal source begins at relative 5PN order, the product $K\,\mathcal H_{\rm tidal}$ starts as $v^4$, not as a negative power of $v$. This means that, within the present adiabatic 5PN tidal model, the phase-transition contribution is weighted toward the later and stronger-field portion of the inspiral rather than toward the earliest weak-field stage.

For the sharp first-order transition, the leading-order dephasing follows immediately from Eqs.~\eqref{eq:DeltaPhi_step_general_section4}, \eqref{eq:K_pn_section4}, and \eqref{eq:Htidal_repeated_section4}. Retaining only the lowest term in each expansion gives
\begin{equation}
K_{\rm LO}(v)=\frac{5}{32q}v^{-6},
\qquad
\mathcal H_{\rm tidal}^{\rm LO}(v)=\kappa_0 v^{10},
\end{equation}
and therefore
\begin{equation}
\Delta\Phi_{\rm step}^{\rm LO}
=
-\frac{\kappa_0\Delta\Lambda}{32q}
\left(
v_{\rm ISCO}^5-v_c^5
\right).
\label{eq:DeltaPhi_step_LO_section4}
\end{equation}

This expression already captures the two leading physical dependencies. Since $\Delta\Lambda<0$ for a transition to a denser quark core, it follows that $\Delta\Phi_{\rm step}^{\rm LO}>0$: tidal dissipation is reduced after the transition, and the inspiral therefore accumulates \emph{more} orbital phase than the purely hadronic baseline before reaching the same final orbital state. Moreover, the result is highly sensitive to both the transition location and the Kerr strong-field endpoint, through $v_c$ and $v_{\rm ISCO}(\hat a)$.

Going beyond leading order, the $v^9$-truncated step-transition result may be written schematically as
\begin{align}
\Delta\Phi_{\rm step}^{(v^9)}
&=
-\Delta\Lambda
\int_{v_c}^{v_{\rm ISCO}}
K^{(v^{-1})}(v,\hat a)\,
\mathcal H_{\rm tidal}^{(v^4)}(v,\hat a)\,dv
\nonumber\\
&=
-\frac{\kappa_0\Delta\Lambda}{32q}
\left(v_{\rm ISCO}^5-v_c^5\right)
-\frac{\kappa_0\Delta\Lambda}{7q}
\left(
\frac{5h_2}{32}+\frac{3715}{10752}
\right)
\left(v_{\rm ISCO}^7-v_c^7\right)
\nonumber\\
&\quad
-\frac{\kappa_0\Delta\Lambda}{8q}
\left(
\frac{5}{32}h_{3s}\hat a
-\pi
+\frac{565}{48}\hat a
\right)
\left(v_{\rm ISCO}^8-v_c^8\right)
+\cdots ,
\label{eq:DeltaPhi_step_v9_section4}
\end{align}

where only the first few terms are shown explicitly, and the full closed-form expression is presented in Appendix~\ref{app:pn_kernel_pt}. Equation~\eqref{eq:DeltaPhi_step_v9_section4} makes the role of spin especially transparent: already at the order displayed here, $\hat a$ affects the dephasing both indirectly through the endpoint $v_{\rm ISCO}(\hat a)$ and directly through the coefficients multiplying the higher-order terms.

The mixed-phase case admits an analogous decomposition. At leading order,
\begin{align}
\Delta\Phi_{\rm mix}^{\rm LO}
&=
-\frac{5\kappa_0(\Lambda_Q-\Lambda_H)}{32q}
\left[
\int_{v_h}^{v_q}v^4
S\!\left(\frac{v-v_h}{v_q-v_h}\right)\,dv
+
\int_{v_q}^{v_{\rm ISCO}}v^4\,dv
\right]
\nonumber\\
&=
-\frac{5\kappa_0(\Lambda_Q-\Lambda_H)}{32q}
\left[
I_{\rm mix}
+\frac{v_{\rm ISCO}^5-v_q^5}{5}
\right],
\label{eq:DeltaPhi_mix_LO_section4}
\end{align}
where the transition-region integral
\begin{equation}
I_{\rm mix}
\equiv
\int_{v_h}^{v_q}v^4
S\!\left(\frac{v-v_h}{v_q-v_h}\right)\,dv
\label{eq:Imix_def_section4}
\end{equation}
is an explicit polynomial in $(v_h,v_q)$ because $S(y)$ is cubic. Thus the finite-width transition preserves analytic control while smoothing the onset of the signal. Physically, the effect of the mixed phase is to spread the dephasing source over a finite interval in orbital space; mathematically, it replaces the single lower integration limit $v_c$ by a competition between a weighted transition-region contribution and a tail identical in structure to the sharp-transition result.

A final point deserves emphasis. The formulas above show that the observable phase shift is controlled by three distinct layers of physics. The first is the universal Kerr geometry, encoded in $K(v,\hat a)$. The second is the microphysical magnitude of the transition, encoded in $\Delta\Lambda$ or $(\Lambda_Q-\Lambda_H)$. The third is the orbital localization of the transition, encoded in $v_c$ or $(v_h,v_q)$. This factorization is precisely what makes the present framework analytically useful: it separates what is fixed by general relativity from what is supplied by dense-matter physics and from what is determined by the dynamical history of the inspiral. In Sec.~\ref{sec:numerical} we exploit this structure to generate parameter scans, contour plots, and time-domain waveforms directly from the closed expressions derived here.

\section{\label{sec:numerical}Numerical Results and Waveform Generation}

In this section we translate the analytical results of Sec.~\ref{sec:derivation} into observable diagnostics. Our goal is not yet a full Bayesian forecast or a detector-level parameter-estimation study, but rather a controlled demonstration of how the phase-transition source term propagates through the Kerr dephasing kernel into waveform-level signatures. The numerical results therefore serve three complementary purposes. First, they display the microphysical input in a transparent way. Second, they validate the analytical scaling laws across the relevant parameter space. Third, they show explicitly how the accumulated phase shift appears in the time-domain gravitational waveform.

The numerical implementation follows directly from the formalism developed in the previous sections. The background inspiral is described by the Kerr point-particle kernel $K(v,\hat a)$, the microphysical transition enters through the effective tidal response function $\delta\Lambda(v)$, and the total phase shift is determined by
\begin{equation}
\Delta\Phi_{\rm PT}
=
-\int
K(v,\hat a)\,
\delta\Lambda(v)\,
\mathcal H_{\rm tidal}(v,\hat a)\,dv.
\label{eq:DeltaPhi_master_sec5}
\end{equation}
The three figures discussed below should therefore be viewed as different projections of the same analytic framework: Fig.~\ref{fig:lambda_transition} displays the microphysical response profile itself, Fig.~\ref{fig:param_scan} shows how the closed-form dephasing formula behaves across the $(v_c,\hat a)$ plane, and Fig.~\ref{fig:waveform_dephasing} illustrates the corresponding waveform-level phase drift.

\subsection{Parameter Space and Analytical Validation}

\begin{figure}[htbp]
\centering
\includegraphics[width=0.92\columnwidth]{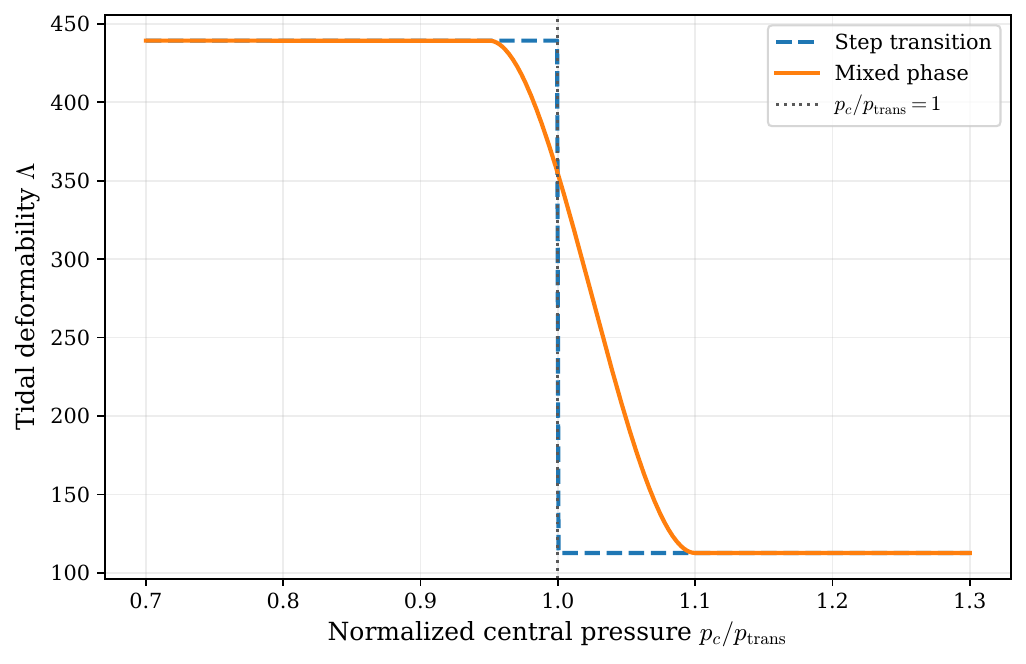}
\caption{\label{fig:lambda_transition}
Dimensionless tidal deformability $\Lambda$ as a function of the normalized central pressure $p_c/p_{\rm trans}$. The dashed curve shows an idealized sharp first-order transition, while the solid curve shows a finite-width mixed-phase transition. The reduction from the hadronic branch to the quark-core branch defines the effective microphysical input $\delta\Lambda$ that enters the dephasing integral.}
\end{figure}

We begin with the microphysical input. Figure~\ref{fig:lambda_transition} shows the dimensionless tidal deformability $\Lambda$ as a function of the normalized central pressure $p_c/p_{\rm trans}$. Two idealized transition patterns are displayed. The dashed curve represents the limiting case of a sharp first-order transition, while the solid curve represents a finite-width mixed phase. In the effective inspiral description, this figure is the direct precursor of the orbital response function $\delta\Lambda(v)$ introduced in Sec.~\ref{sec:qcd_model}. The sharp profile corresponds to the step model of Eq.~\eqref{eq:Lambda_step_qcdmodel}, whereas the smooth profile corresponds to the mixed-phase interpolation of Eq.~\eqref{eq:Lambda_mixed_qcdmodel}.

The physical content of Fig.~\ref{fig:lambda_transition} is straightforward but fundamental. The hadronic configuration is more easily deformable, while the quark-core configuration is more compact and therefore less responsive to the external tidal field. Consequently,
\begin{equation}
\Delta\Lambda=\Lambda_Q-\Lambda_H<0.
\end{equation}
This sign has an immediate dynamical interpretation. Once the phase transition occurs, the tidal contribution to the total dissipation is weakened relative to the purely hadronic baseline. The phase-transition signal is therefore not an amplitude feature in the usual sense, but a cumulative timing effect that builds up through the orbital phase evolution.

\begin{figure}[htbp]
\centering
\includegraphics[width=0.95\columnwidth]{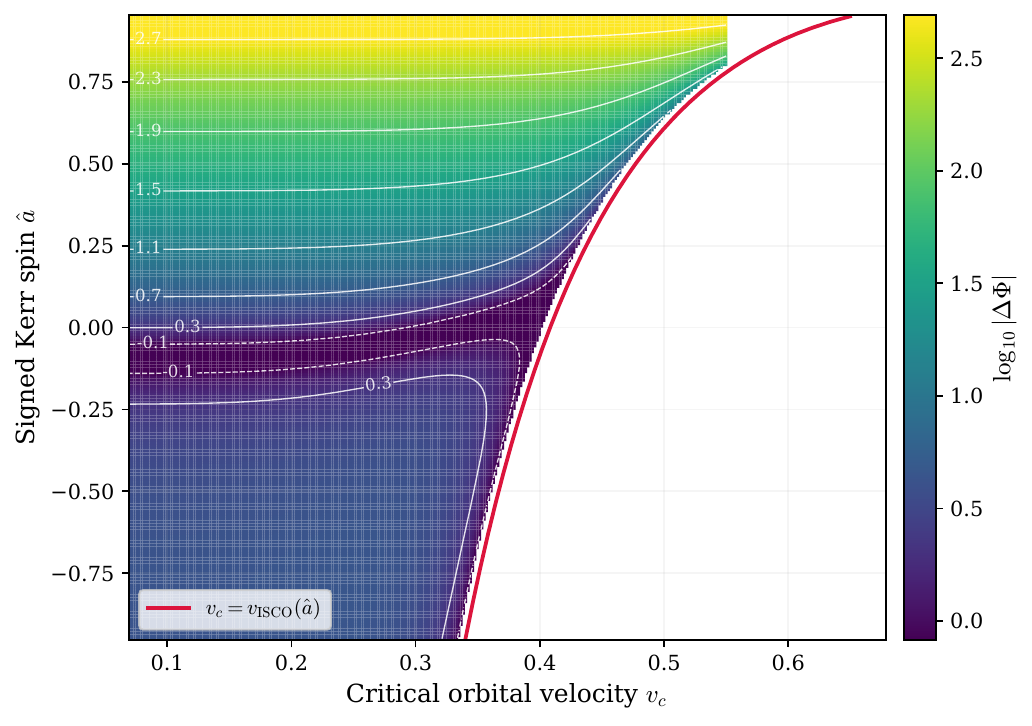}
\caption{\label{fig:param_scan}
Contour plot of the accumulated dephasing $\log_{10}|\Delta\Phi|$ in the $(v_c,\hat a)$ plane, computed from the $v^9$-truncated analytical formula. The red curve denotes the boundary $v_c=v_{\rm ISCO}(\hat a)$, beyond which the transition would occur only at or after plunge.}
\end{figure}

The next step is to insert this microphysical input into the analytical dephasing formula. Figure~\ref{fig:param_scan} shows the resulting contour map of $\log_{10}|\Delta\Phi|$ in the parameter plane spanned by the critical orbital velocity $v_c$ and the signed Kerr spin $\hat a$, using the $v^9$-truncated analytical expression derived in Sec.~\ref{sec:derivation}. The red curve marks the kinematic boundary
\begin{equation}
v_c=v_{\rm ISCO}(\hat a),
\label{eq:isco_boundary_sec5}
\end{equation}
beyond which no post-transition inspiral remains. The physically admissible region therefore lies to the left of the boundary.

Several important trends are immediately visible in Fig.~\ref{fig:param_scan}. First, for fixed spin, the dephasing grows as the transition occurs earlier in orbital space, that is, as $v_c$ moves away from the plunge boundary. Second, for fixed $v_c$, the phase shift is enhanced for positive $\hat a$, corresponding to prograde configurations. This behavior is fully consistent with the analytical structure of the problem. Positive spin moves the ISCO inward and increases the strong-field interval over which the post-transition inspiral can accumulate phase, while negative spin shifts the ISCO outward and suppresses the available interval. The contour pattern therefore gives a direct graphical realization of the factorization
\begin{equation}
\Delta\Phi_{\rm PT}
\sim
\big[\text{Kerr kernel}\big]
\times
\big[\text{microphysical jump}\big]
\times
\big[\text{post-transition inspiral interval}\big].
\end{equation}

The same figure also provides a useful validation of the analytical scaling law. At leading order one has
\begin{equation}
\Delta\Phi_{\rm step}^{\rm LO}
=
-\frac{\kappa_0\Delta\Lambda}{32q}
\left(v_{\rm ISCO}^5-v_c^5\right),
\label{eq:LO_validation_sec5}
\end{equation}
which predicts a monotonic dependence on both $v_c$ and $v_{\rm ISCO}(\hat a)$. The contour map follows precisely this logic, while the higher-order $v^7$, $v^8$, and $v^9$ terms introduce the nontrivial distortions and spin-dependent structure seen in the full result. In this sense, Fig.~\ref{fig:param_scan} is not merely illustrative; it is a compact validation of the symbolic analytical pipeline itself.

\subsection{Time-domain Gravitational Waveforms}

We now turn to the time domain, where the accumulated dephasing becomes directly visible as a mismatch between two waveform histories. At leading adiabatic order, the plus polarization may be written schematically as
\begin{equation}
h_+(t)=A(t)\cos\Phi(t),
\label{eq:hplus_baseline_sec5}
\end{equation}
with $A(t)$ a slowly varying amplitude envelope and $\Phi(t)$ the accumulated phase. In the presence of a phase transition, this becomes
\begin{equation}
h_+^{\rm PT}(t)=A(t)\cos\!\left[\Phi_0(t)+\Delta\Phi_{\rm PT}(t)\right],
\label{eq:hplus_pt_sec5}
\end{equation}
where $\Phi_0(t)$ denotes the purely hadronic baseline and $\Delta\Phi_{\rm PT}(t)=\Delta\Phi_{\rm PT}(v(t))$ is the cumulative correction generated by the transition.

\begin{figure*}[htbp]
\centering
\includegraphics[width=0.96\textwidth]{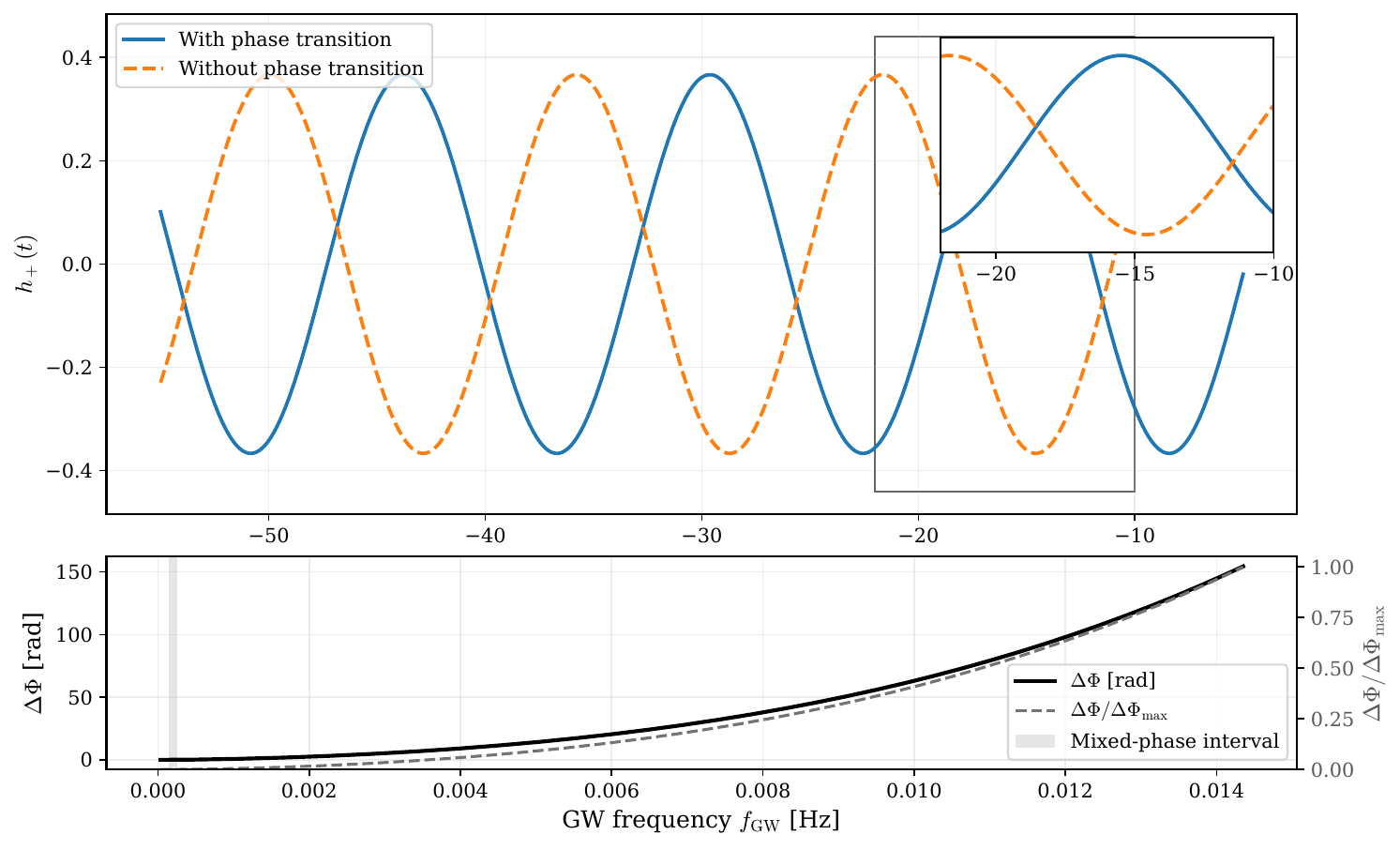}
\caption{\label{fig:waveform_dephasing}
Waveform-level manifestation of the phase-transition-induced dephasing. The main waveform panel compares the baseline signal with the phase-transition-modified signal during the late inspiral. The accompanying phase-accumulation panels show the growth of $\Delta\Phi$ with gravitational-wave frequency and the corresponding normalized cumulative profile, making explicit where in the inspiral the microphysical signal is accumulated.}
\end{figure*}

Figure~\ref{fig:waveform_dephasing} presents the waveform-level manifestation of this effect. The main panel compares the late-inspiral baseline waveform with the waveform modified by the phase transition. The additional panels show the cumulative dephasing as a function of gravitational-wave frequency and the corresponding normalized growth profile. Together they make clear that the phase transition acts primarily through a secular phase drift rather than through an instantaneous amplitude distortion.

The main waveform comparison in Fig.~\ref{fig:waveform_dephasing} shows that the two histories remain nearly indistinguishable at earlier times, when the integrated correction is still small, but drift progressively out of phase as the inspiral approaches the relativistic end state. This is exactly what the formalism predicts. Because the phase transition enters through the dissipation sector, it changes the orbital timing and therefore the accumulated waveform phase, rather than producing a sudden burst-like feature.

The phase-accumulation panel in Fig.~\ref{fig:waveform_dephasing} makes the same statement in a more transparent way. At low frequencies the phase shift is negligible, reflecting the weakness of the tidal sector in the early inspiral. As the orbit enters the mixed-phase interval, the phase shift begins to build rapidly, and after the transition is complete it continues to grow through the remaining strong-field portion of the inspiral. This is the expected consequence of the leading-order scaling
\begin{equation}
K(v,\hat a)\,\mathcal H_{\rm tidal}(v,\hat a)\sim v^4,
\end{equation}
which weights the phase-transition source toward the later, relativistic regime rather than toward the earliest weak-field stage.

The normalized growth profile shown in Fig.~\ref{fig:waveform_dephasing} is useful for a complementary reason. By factoring out the absolute scale of the phase shift, it isolates the temporal structure of the signal accumulation. One sees immediately that the transition does not contribute uniformly across the inspiral. Instead, the dominant fraction of the measurable phase offset is accumulated after the system has already entered the high-frequency regime. This observation reinforces the lesson of Fig.~\ref{fig:param_scan}: the strong-field geometry of the Kerr background is not a passive backdrop, but an active amplifier of the dense-matter imprint.

Taken together, Figs.~\ref{fig:lambda_transition}--\ref{fig:waveform_dephasing} form a coherent chain of interpretation. Figure~\ref{fig:lambda_transition} defines the microphysical response model. Figure~\ref{fig:param_scan} shows how that response is amplified or suppressed across the $(v_c,\hat a)$ parameter space. Figure~\ref{fig:waveform_dephasing} then demonstrates how the same mechanism appears in the observable time-domain signal. The three figures therefore play distinct but complementary roles: microphysical characterization, analytical validation, and waveform interpretation. In that sense, they close the logical loop between the dense-matter model of Sec.~\ref{sec:qcd_model}, the dephasing formalism of Sec.~\ref{sec:derivation}, and the gravitational-wave phenomenology relevant to future space-based observations.

\section{\label{sec:conclusion}Conclusion}

In this work we have developed an analytical framework for describing long-term gravitational-wave dephasing induced by a dense-matter phase transition inside the neutron-star secondary of an extreme-mass-ratio inspiral in Kerr spacetime. The central idea of the paper is conceptually simple but physically nontrivial: the relativistic orbital motion is governed by the background Kerr geometry, while the nonperturbative microphysics of dense QCD matter enters the waveform only through the tidal response of the secondary. This separation of scales makes it possible to encode the hadron--quark transition as a controlled perturbation of the dissipative sector, without sacrificing the strong-field structure of the Kerr inspiral itself.

The analysis proceeded in three steps. First, we formulated the circular--equatorial EMRI dynamics using the exact Kerr orbital energy and a Teukolsky-inspired post-Newtonian expansion of the point-particle flux. This allowed us to define a background phase kernel $K(v,\hat a)$ that isolates the purely relativistic contribution of the Kerr spacetime. Second, we modeled the dense-matter transition through the effective tidal deformability $\Lambda$, introducing both an idealized sharp first-order transition and a finite-width mixed-phase interpolation. In this way the microphysical problem was reduced to the behavior of a single orbital response function $\delta\Lambda(v)$. Third, by inserting this response into the energy-balance law, we derived a closed dephasing functional, Eq.~\eqref{eq:master_dephasing_section4}, which expresses the total phase shift as the convolution of the Kerr background kernel with the transition-induced tidal source.

At the analytical level, the most important outcome of the paper is the explicit scaling structure of the phase shift. For a sharp transition, the leading contribution takes the form
\begin{equation}
\Delta\Phi_{\rm step}^{\rm LO}
=
-\frac{\kappa_0\Delta\Lambda}{32q}
\left(v_{\rm ISCO}^5-v_c^5\right),
\end{equation}
while the higher-order $v^7$, $v^8$, and $v^9$ corrections introduce direct spin dependence as well as nontrivial post-Newtonian distortions. This result makes several physical facts manifest. First, for the phenomenologically relevant case $\Delta\Lambda<0$, the accumulated phase shift is positive: after the transition the star becomes less deformable, the tidal contribution to the total dissipation is reduced, and the system accumulates more orbital phase than the purely hadronic baseline before reaching the same final orbital state. Second, the phase shift is highly sensitive to the location of the transition in orbital space. Third, the black-hole spin acts as a strong-field amplifier by modifying both the Kerr phase kernel and the available post-transition inspiral interval through the spin dependence of $v_{\rm ISCO}(\hat a)$.

The numerical results confirm and clarify this analytical picture. The transition profiles for $\Lambda$ demonstrate explicitly how a sharp first-order jump and a mixed-phase transition define distinct source functions in orbital space. The parameter-space scan in the $(v_c,\hat a)$ plane shows that the accumulated dephasing is strongly enhanced for earlier transitions and for prograde configurations, precisely as predicted by the closed analytic formulas. The waveform-level comparison further shows that the effect appears not as a sudden amplitude anomaly but as a coherent late-inspiral phase drift. This is the physically important point: the dense-matter signal is encoded as a cumulative timing effect, and EMRIs are exactly the class of sources for which such timing effects can become observable because of the enormous number of gravitational-wave cycles accumulated in band.

The broader significance of this framework is methodological as much as phenomenological. By expressing the phase-transition effect entirely through the map
\begin{equation}
\delta\Lambda(v)
\longrightarrow
\Delta\Phi_{\rm PT},
\end{equation}
we have separated the problem into a universal relativistic sector and a model-dependent microphysical sector. This factorization is useful for at least two reasons. On the one hand, it allows one to explore a wide class of hybrid-star equations of state without rebuilding the orbital dynamics from scratch. On the other hand, it provides a transparent pathway for future improvements: once more accurate tidal fluxes, self-force information, or numerical Teukolsky amplitudes become available, they can be inserted into the same formal structure without changing the conceptual organization of the calculation.

At the same time, the scope of the present work should be stated clearly. We have restricted attention to circular equatorial inspirals in the adiabatic regime and to perturbative tidal corrections. We have not included eccentricity, inclination, spin precession of the secondary, nonlinear hydrodynamical conversion fronts, nor resonant interface or composition modes. We have also not attempted a full detector-response study or a Bayesian parameter-estimation analysis. These omissions are deliberate. The purpose of the present paper has been to establish the analytical backbone of the problem and to show, in the cleanest possible setting, how a QCD phase transition can be translated into a controlled waveform-level dephasing signal.

Several natural extensions follow immediately from the present formalism. The first is the incorporation of more accurate Teukolsky-based tidal fluxes and higher-order Kerr spin couplings, replacing the present phenomenological source coefficients by fully calibrated strong-field amplitudes. The second is the inclusion of generic Kerr orbits, especially eccentric inspirals, where the transition may couple to a richer spectrum of orbital harmonics and possibly to transient resonant effects. The third is the integration of this framework into LISA data-analysis pipelines, where one can test whether the dephasing derived here remains distinguishable once detector noise, parameter covariances, and waveform systematics are taken into account. Finally, a particularly promising direction is to connect the effective parameters $(\Lambda_H,\Lambda_Q,v_c,\Delta v)$ to explicit hybrid-star sequences derived from microscopic dense-matter models, thereby turning the present formalism into a practical interface between gravitational-wave astronomy and high-density QCD phenomenology.

In summary, we have shown that a dense-matter phase transition inside the neutron-star secondary of a Kerr EMRI can be encoded analytically as a perturbation of the tidal response and propagated consistently into the gravitational-wave phase. The result is a controlled strong-field dephasing formalism that captures the interplay between Kerr geometry, long-term inspiral memory, and QCD-motivated changes in stellar structure. This provides a new analytical route for using future space-based gravitational-wave observations to probe matter under conditions inaccessible to terrestrial experiments. In that sense, the present work suggests that EMRIs may eventually serve not only as laboratories of strong gravity, but also as precision probes of the phase structure of dense QCD matter.

\begin{acknowledgments}
The first author would like to express his sincere gratitude to Liangyu Luo for valuable help with the derivation of the physical formulas and the analysis of the data, as well as for continuous support and encouragement throughout this research. Jie Shi is acknowledged for contributions to part of the calculations and for assistance with the formatting and revision of the manuscript. The authors also acknowledge internal funding support from Lomonosov Moscow State University.
\end{acknowledgments}

\appendix

\section{\label{app:exact_kerr}Exact Kerr orbital quantities}

In this appendix we collect the exact Kerr orbital relations, the post-Newtonian (PN) ingredients used in the analytic phase kernel, and the closed-form dephasing expressions for both the sharp-transition and mixed-phase scenarios. These formulas provide the technical backbone of the analytic results quoted in the main text.

For equatorial circular geodesics in Kerr spacetime, we introduce the signed spin parameter
\begin{equation}
\hat a \equiv \sigma \chi,
\qquad
\sigma=
\begin{cases}
+1, & \text{prograde},\\
-1, & \text{retrograde},
\end{cases}
\end{equation}
where $\chi=a/M$ is the usual dimensionless Kerr spin. The orbital frequency measured at infinity is
\begin{equation}
\Omega
=
\frac{1}{M\left[(r/M)^{3/2}+\hat a\right]},
\label{eq:Omega_exact_app}
\end{equation}
and we define the invariant PN velocity parameter
\begin{equation}
v\equiv (M\Omega)^{1/3}.
\label{eq:v_def_app}
\end{equation}
In terms of $v$, the inverse-radius variable becomes
\begin{equation}
u(v,\hat a)
=
\frac{v^2}{\left(1-\hat a v^3\right)^{2/3}}
=
\frac{M}{r}.
\label{eq:u_of_v_app}
\end{equation}

The specific orbital energy $\tilde E\equiv E/\mu$ for circular equatorial motion is
\begin{equation}
\tilde E(v,\hat a)
=
\frac{1-2u+\hat a\,u^{3/2}}
{\sqrt{1-3u+2\hat a\,u^{3/2}}},
\label{eq:Etilde_exact_app}
\end{equation}
which differentiates to
\begin{equation}
\frac{d\tilde E}{dv}
=
\frac{
v\left(3\hat a^2u^2-8\hat a u^{3/2}+6u-1\right)
}{
\left(1-\hat a v^3\right)^{5/3}
\left(1-3u+2\hat a u^{3/2}\right)^{3/2}
}.
\label{eq:dEtilde_exact_app}
\end{equation}

The ISCO radius is given by the standard Bardeen--Press--Teukolsky formula,
\begin{align}
\frac{r_{\rm ISCO}}{M}
&=
3+Z_2-\sigma\sqrt{(3-Z_1)(3+Z_1+2Z_2)},
\label{eq:rISCO_app}
\\
Z_1
&=
1+\left(1-\chi^2\right)^{1/3}
\left[(1+\chi)^{1/3}+(1-\chi)^{1/3}\right],
\\
Z_2
&=
\sqrt{3\chi^2+Z_1^2},
\end{align}
with the associated endpoint frequency parameter
\begin{equation}
v_{\rm ISCO}(\hat a)
=
\left[
\frac{1}{(r_{\rm ISCO}/M)^{3/2}+\hat a}
\right]^{1/3}.
\label{eq:vISCO_app}
\end{equation}

These exact Kerr relations are used to connect the orbital dynamics to the PN-expanded phase kernel introduced below.

\section{\label{app:pn_kernel_pt}PN-expanded phase kernel}

To describe a finite-width hadron--quark transition region, we further define the dimensionless interpolation variable
\begin{equation}
y=\frac{v-v_h}{v_q-v_h},
\label{eq:app_y}
\end{equation}
together with the smoothstep function
\begin{equation}
S(y)=-2y^{3}+3y^{2}.
\label{eq:app_smoothstep}
\end{equation}
By construction, $S(0)=0$, $S(1)=1$, and both the function and its first derivative remain continuous across the boundaries of the mixed-phase interval.

To the order retained in the symbolic pipeline, the PN-expanded specific orbital energy is
\begin{align}
\tilde E_{\rm PN}(v,\hat a)
&=
1
-\frac{v^{2}}{2}
+\frac{3v^{4}}{8}
+\frac{27v^{6}}{16}
+\frac{675v^{8}}{128}
+\frac{3969v^{10}}{256}
\nonumber\\
&\quad
-\frac{4\hat a}{3}v^{5}
-4\hat a v^{7}
-\frac{27\hat a}{2}v^{9},
\label{eq:app_Epn}
\end{align}
which differentiates to
\begin{align}
\frac{d\tilde E_{\rm PN}}{dv}
&=
-v
+\frac{3v^{3}}{2}
+\frac{81v^{5}}{8}
+\frac{675v^{7}}{16}
+\frac{19845v^{9}}{128}
\nonumber\\
&\quad
-\frac{20\hat a}{3}v^{4}
-28\hat a v^{6}
-\frac{243\hat a}{2}v^{8}.
\label{eq:app_dEpn}
\end{align}

The normalized point-particle flux is taken in the form
\begin{align}
\hat{\mathcal F}_{\rm PP}(v,\hat a)
&=
\frac{32v^{10}}{5}
\Bigg[
1
-\frac{1247}{336}v^{2}
+\left(4\pi-\frac{11\hat a}{4}\right)v^{3}
\nonumber\\
&\qquad
+\left(\frac{33\hat a^{2}}{16}-\frac{44711}{9072}\right)v^{4}
+\left(-\frac{59\hat a}{16}-\frac{8191\pi}{672}\right)v^{5}
\Bigg].
\label{eq:app_Fpp}
\end{align}

Combining the PN-expanded orbital energy and flux yields the dephasing kernel
\begin{align}
K_{\rm PN}(v,\hat a)
&=
\frac{5}{32qv^{6}}
+\frac{3715}{10752qv^{4}}
-\frac{5\pi}{8qv^{3}}
+\frac{15293365}{32514048qv^{2}}
-\frac{38645\pi}{21504qv}
\nonumber\\
&\quad
+\hat a
\left(
\frac{565}{384qv^{3}}
+\frac{732985}{64512qv}
\right).
\label{eq:app_Kpn}
\end{align}

The effective tidal source term is parameterized as
\begin{equation}
\mathcal H_{\rm tidal}(v,\hat a)
=
\kappa_{0}v^{10}
\left(
1+h_{2}v^{2}+\hat a\,h_{3s}v^{3}+h_{4}v^{4}
\right).
\label{eq:app_Htidal}
\end{equation}
This form is sufficiently general to encode the leading response amplitude together with the dominant higher-order corrections relevant for the phase-shift calculation.

\section{\label{app:closed_forms}Closed-form phase-shift formulas}

We now present the explicit closed-form dephasing expressions for the two transition patterns studied in the paper: a sharp first-order jump and a finite-width mixed phase. The formulas below are written in a way that makes the dependence on the transition scales and response coefficients fully explicit.

\subsection{Sharp first-order transition}

For a sharp first-order transition, the tidal deformability correction is modeled by a step profile,
\begin{equation}
\delta\Lambda_{\rm step}(v)=\Delta\Lambda\,\Theta(v-v_c),
\qquad
\Delta\Lambda=\Lambda_Q-\Lambda_H,
\label{eq:app_dLambda_step}
\end{equation}
so that the accumulated dephasing is
\begin{equation}
\Delta\Phi_{\rm step}
=
-\Delta\Lambda
\int_{v_c}^{v_{\rm ISCO}}
K(v,\hat a)\,
\mathcal H_{\rm tidal}(v,\hat a)\,dv.
\label{eq:app_DPhi_step_int}
\end{equation}

Using Eqs.~\eqref{eq:app_Kpn} and \eqref{eq:app_Htidal}, the $v^{9}$-truncated integrand takes the form
\begin{align}
I_{\rm step}^{(v^{9})}(v)
&=
-\frac{5\Delta\Lambda\kappa_{0}}{32q}v^{4}
-\frac{3715\Delta\Lambda\kappa_{0}}{10752q}v^{6}
+\frac{5\pi\Delta\Lambda\kappa_{0}}{8q}v^{7}
-\frac{15293365\Delta\Lambda\kappa_{0}}{32514048q}v^{8}
+\frac{38645\pi\Delta\Lambda\kappa_{0}}{21504q}v^{9}
\nonumber\\
&\quad
+h_{2}\left(
-\frac{5\Delta\Lambda\kappa_{0}}{32q}v^{6}
-\frac{3715\Delta\Lambda\kappa_{0}}{10752q}v^{8}
+\frac{5\pi\Delta\Lambda\kappa_{0}}{8q}v^{9}
\right)
\nonumber\\
&\quad
+h_{4}\left(
-\frac{5\Delta\Lambda\kappa_{0}}{32q}v^{8}
\right)
\nonumber\\
&\quad
+\hat a\Bigg[
-\frac{5\Delta\Lambda h_{3s}\kappa_{0}}{32q}v^{7}
-\frac{565\Delta\Lambda\kappa_{0}}{384q}v^{7}
-\frac{3715\Delta\Lambda h_{3s}\kappa_{0}}{10752q}v^{9}
-\frac{565\Delta\Lambda h_{2}\kappa_{0}}{384q}v^{9}
-\frac{732985\Delta\Lambda\kappa_{0}}{64512q}v^{9}
\Bigg].
\label{eq:app_Istep_v9}
\end{align}

Integrating term by term gives
\begin{align}
\Delta\Phi_{\rm step}^{(v^{9})}
&=
-\frac{\Delta\Lambda\kappa_{0}}{32q}
\left(v_{\rm ISCO}^{5}-v_c^{5}\right)
-\frac{3715\Delta\Lambda\kappa_{0}}{75264q}
\left(v_{\rm ISCO}^{7}-v_c^{7}\right)
+\frac{5\pi\Delta\Lambda\kappa_{0}}{64q}
\left(v_{\rm ISCO}^{8}-v_c^{8}\right)
\nonumber\\
&\quad
-\frac{15293365\Delta\Lambda\kappa_{0}}{292626432q}
\left(v_{\rm ISCO}^{9}-v_c^{9}\right)
+\frac{38645\pi\Delta\Lambda\kappa_{0}}{215040q}
\left(v_{\rm ISCO}^{10}-v_c^{10}\right)
\nonumber\\
&\quad
+h_{2}\Bigg[
-\frac{5\Delta\Lambda\kappa_{0}}{224q}
\left(v_{\rm ISCO}^{7}-v_c^{7}\right)
-\frac{3715\Delta\Lambda\kappa_{0}}{96768q}
\left(v_{\rm ISCO}^{9}-v_c^{9}\right)
+\frac{\pi\Delta\Lambda\kappa_{0}}{16q}
\left(v_{\rm ISCO}^{10}-v_c^{10}\right)
\Bigg]
\nonumber\\
&\quad
+h_{4}\Bigg[
-\frac{5\Delta\Lambda\kappa_{0}}{288q}
\left(v_{\rm ISCO}^{9}-v_c^{9}\right)
\Bigg]
\nonumber\\
&\quad
+\hat a\Bigg[
-\frac{5\Delta\Lambda h_{3s}\kappa_{0}}{256q}
\left(v_{\rm ISCO}^{8}-v_c^{8}\right)
-\frac{565\Delta\Lambda\kappa_{0}}{3072q}
\left(v_{\rm ISCO}^{8}-v_c^{8}\right)
\nonumber\\
&\qquad\qquad
-\frac{3715\Delta\Lambda h_{3s}\kappa_{0}}{107520q}
\left(v_{\rm ISCO}^{10}-v_c^{10}\right)
-\frac{565\Delta\Lambda h_{2}\kappa_{0}}{3840q}
\left(v_{\rm ISCO}^{10}-v_c^{10}\right)
-\frac{732985\Delta\Lambda\kappa_{0}}{645120q}
\left(v_{\rm ISCO}^{10}-v_c^{10}\right)
\Bigg].
\label{eq:app_DPhi_step_v9}
\end{align}

At leading order, the kernel reduces to
\begin{equation}
K_{\rm LO}(v)=\frac{5}{32qv^{6}},
\label{eq:app_KLO}
\end{equation}
and therefore
\begin{equation}
\Delta\Phi_{\rm step}^{\rm LO}
=
\frac{\Delta\Lambda\kappa_{0}}{32q}
\left(-v_{\rm ISCO}^{5}+v_c^{5}\right).
\label{eq:app_DPhi_step_LO}
\end{equation}
This leading-order expression already captures the basic scaling of the dephasing with the transition jump and the cutoff frequency.

\subsection{Finite-width mixed phase}

For the mixed-phase model we define
\begin{equation}
\Delta\Lambda_{HQ}\equiv \Lambda_Q-\Lambda_H,
\label{eq:app_DLambdaHQ}
\end{equation}
and use the interpolating profile
\begin{equation}
\delta\Lambda_{\rm mix}(v)=
\begin{cases}
0, & v\le v_h,\\[4pt]
\Delta\Lambda_{HQ}\,S(y), & v_h<v<v_q,\\[4pt]
\Delta\Lambda_{HQ}, & v\ge v_q.
\end{cases}
\label{eq:app_dLambda_mix}
\end{equation}
The resulting phase shift naturally separates into a contribution from the interpolation region and a contribution from the fully developed quark-phase tail,
\begin{equation}
\Delta\Phi_{\rm mix}^{(v^{9})}
=
\int_{v_h}^{v_q} I_{\rm mix,mid}^{(v^{9})}(v)\,dv
+
\int_{v_q}^{v_{\rm ISCO}} I_{\rm mix,tail}^{(v^{9})}(v)\,dv.
\label{eq:app_DPhi_mix_split}
\end{equation}

The tail integrand is
\begin{align}
I_{\rm mix,tail}^{(v^{9})}(v)
&=
-\frac{5\Delta\Lambda_{HQ}\kappa_{0}}{32q}v^{4}
-\frac{3715\Delta\Lambda_{HQ}\kappa_{0}}{10752q}v^{6}
+\frac{5\pi\Delta\Lambda_{HQ}\kappa_{0}}{8q}v^{7}
-\frac{15293365\Delta\Lambda_{HQ}\kappa_{0}}{32514048q}v^{8}
+\frac{38645\pi\Delta\Lambda_{HQ}\kappa_{0}}{21504q}v^{9}
\nonumber\\
&\quad
+h_{2}\Bigl(
-\frac{5\Delta\Lambda_{HQ}\kappa_{0}}{32q}v^{6}
-\frac{3715\Delta\Lambda_{HQ}\kappa_{0}}{10752q}v^{8}
+\frac{5\pi\Delta\Lambda_{HQ}\kappa_{0}}{8q}v^{9}
\Bigr)
\nonumber\\
&\quad
+h_{4}\Bigl(
-\frac{5\Delta\Lambda_{HQ}\kappa_{0}}{32q}v^{8}
\Bigr)
\nonumber\\
&\quad
+\hat a\Biggl[
-\frac{5\Delta\Lambda_{HQ} h_{3s}\kappa_{0}}{32q}v^{7}
-\frac{565\Delta\Lambda_{HQ}\kappa_{0}}{384q}v^{7}
\nonumber\\
&\qquad\qquad
-\frac{3715\Delta\Lambda_{HQ} h_{3s}\kappa_{0}}{10752q}v^{9}
-\frac{565\Delta\Lambda_{HQ} h_{2}\kappa_{0}}{384q}v^{9}
-\frac{732985\Delta\Lambda_{HQ}\kappa_{0}}{64512q}v^{9}
\Biggr].
\label{eq:app_Imix_tail}
\end{align}

The middle-interval integrand, obtained after inserting the smoothstep interpolation into the PN-expanded kernel, is
\begingroup
\scriptsize
\setlength{\jot}{0.7pt}
\begin{equation}
\begin{aligned}
\mathcal{I}_{\mathrm{mix,mid}}^{(v^9)}(v)
&= \frac{38645 \pi \Delta\Lambda_{\mathrm{HQ}} \kappa_0 v^{9} v_{\mathrm{h}}^{3}}{- 10752 q v_{\mathrm{h}}^{3} + 32256 q v_{\mathrm{h}}^{2} v_{\mathrm{q}} - 32256 q v_{\mathrm{h}} v_{\mathrm{q}}^{2} + 10752 q v_{\mathrm{q}}^{3}}
+ \frac{15293365 \Delta\Lambda_{\mathrm{HQ}} \kappa_0 v^{9} v_{\mathrm{h}}^{2}}{- 5419008 q v_{\mathrm{h}}^{3} + 16257024 q v_{\mathrm{h}}^{2} v_{\mathrm{q}} - 16257024 q v_{\mathrm{h}} v_{\mathrm{q}}^{2} + 5419008 q v_{\mathrm{q}}^{3}}
\\
&\quad + \frac{38645 \pi \Delta\Lambda_{\mathrm{HQ}} \kappa_0 v^{9} v_{\mathrm{h}}^{2}}{7168 q v_{\mathrm{h}}^{2} - 14336 q v_{\mathrm{h}} v_{\mathrm{q}} + 7168 q v_{\mathrm{q}}^{2}}
+ \frac{15 \pi \Delta\Lambda_{\mathrm{HQ}} \kappa_0 v^{9} v_{\mathrm{h}}}{- 4 q v_{\mathrm{h}}^{3} + 12 q v_{\mathrm{h}}^{2} v_{\mathrm{q}} - 12 q v_{\mathrm{h}} v_{\mathrm{q}}^{2} + 4 q v_{\mathrm{q}}^{3}}
\\
&\quad + \frac{15293365 \Delta\Lambda_{\mathrm{HQ}} \kappa_0 v^{9} v_{\mathrm{h}}}{5419008 q v_{\mathrm{h}}^{2} - 10838016 q v_{\mathrm{h}} v_{\mathrm{q}} + 5419008 q v_{\mathrm{q}}^{2}}
+ \frac{3715 \Delta\Lambda_{\mathrm{HQ}} \kappa_0 v^{9}}{- 5376 q v_{\mathrm{h}}^{3} + 16128 q v_{\mathrm{h}}^{2} v_{\mathrm{q}} - 16128 q v_{\mathrm{h}} v_{\mathrm{q}}^{2} + 5376 q v_{\mathrm{q}}^{3}}
\\
&\quad + \frac{15 \pi \Delta\Lambda_{\mathrm{HQ}} \kappa_0 v^{9}}{8 q v_{\mathrm{h}}^{2} - 16 q v_{\mathrm{h}} v_{\mathrm{q}} + 8 q v_{\mathrm{q}}^{2}}
- \frac{15293365 \Delta\Lambda_{\mathrm{HQ}} \kappa_0 v^{8} v_{\mathrm{h}}^{3}}{- 16257024 q v_{\mathrm{h}}^{3} + 48771072 q v_{\mathrm{h}}^{2} v_{\mathrm{q}} - 48771072 q v_{\mathrm{h}} v_{\mathrm{q}}^{2} + 16257024 q v_{\mathrm{q}}^{3}}
\\
&\quad - \frac{15 \pi \Delta\Lambda_{\mathrm{HQ}} \kappa_0 v^{8} v_{\mathrm{h}}^{2}}{- 4 q v_{\mathrm{h}}^{3} + 12 q v_{\mathrm{h}}^{2} v_{\mathrm{q}} - 12 q v_{\mathrm{h}} v_{\mathrm{q}}^{2} + 4 q v_{\mathrm{q}}^{3}}
- \frac{15293365 \Delta\Lambda_{\mathrm{HQ}} \kappa_0 v^{8} v_{\mathrm{h}}^{2}}{10838016 q v_{\mathrm{h}}^{2} - 21676032 q v_{\mathrm{h}} v_{\mathrm{q}} + 10838016 q v_{\mathrm{q}}^{2}}
\\
&\quad - \frac{3715 \Delta\Lambda_{\mathrm{HQ}} \kappa_0 v^{8} v_{\mathrm{h}}}{- 1792 q v_{\mathrm{h}}^{3} + 5376 q v_{\mathrm{h}}^{2} v_{\mathrm{q}} - 5376 q v_{\mathrm{h}} v_{\mathrm{q}}^{2} + 1792 q v_{\mathrm{q}}^{3}}
- \frac{15 \pi \Delta\Lambda_{\mathrm{HQ}} \kappa_0 v^{8} v_{\mathrm{h}}}{4 q v_{\mathrm{h}}^{2} - 8 q v_{\mathrm{h}} v_{\mathrm{q}} + 4 q v_{\mathrm{q}}^{2}}
\\
&\quad - \frac{3715 \Delta\Lambda_{\mathrm{HQ}} \kappa_0 v^{8}}{3584 q v_{\mathrm{h}}^{2} - 7168 q v_{\mathrm{h}} v_{\mathrm{q}} + 3584 q v_{\mathrm{q}}^{2}}
+ \frac{5 \pi \Delta\Lambda_{\mathrm{HQ}} \kappa_0 v^{7} v_{\mathrm{h}}^{3}}{- 4 q v_{\mathrm{h}}^{3} + 12 q v_{\mathrm{h}}^{2} v_{\mathrm{q}} - 12 q v_{\mathrm{h}} v_{\mathrm{q}}^{2} + 4 q v_{\mathrm{q}}^{3}}
\\
&\quad + \frac{3715 \Delta\Lambda_{\mathrm{HQ}} \kappa_0 v^{7} v_{\mathrm{h}}^{2}}{- 1792 q v_{\mathrm{h}}^{3} + 5376 q v_{\mathrm{h}}^{2} v_{\mathrm{q}} - 5376 q v_{\mathrm{h}} v_{\mathrm{q}}^{2} + 1792 q v_{\mathrm{q}}^{3}}
+ \frac{15 \pi \Delta\Lambda_{\mathrm{HQ}} \kappa_0 v^{7} v_{\mathrm{h}}^{2}}{8 q v_{\mathrm{h}}^{2} - 16 q v_{\mathrm{h}} v_{\mathrm{q}} + 8 q v_{\mathrm{q}}^{2}}
\\
&\quad + \frac{3715 \Delta\Lambda_{\mathrm{HQ}} \kappa_0 v^{7} v_{\mathrm{h}}}{1792 q v_{\mathrm{h}}^{2} - 3584 q v_{\mathrm{h}} v_{\mathrm{q}} + 1792 q v_{\mathrm{q}}^{2}}
+ \frac{5 \Delta\Lambda_{\mathrm{HQ}} \kappa_0 v^{7}}{- 16 q v_{\mathrm{h}}^{3} + 48 q v_{\mathrm{h}}^{2} v_{\mathrm{q}} - 48 q v_{\mathrm{h}} v_{\mathrm{q}}^{2} + 16 q v_{\mathrm{q}}^{3}}
\\
&\quad S- \frac{3715 \Delta\Lambda_{\mathrm{HQ}} \kappa_0 v^{6} v_{\mathrm{h}}^{3}}{- 5376 q v_{\mathrm{h}}^{3} + 16128 q v_{\mathrm{h}}^{2} v_{\mathrm{q}} - 16128 q v_{\mathrm{h}} v_{\mathrm{q}}^{2} + 5376 q v_{\mathrm{q}}^{3}}
- \frac{3715 \Delta\Lambda_{\mathrm{HQ}} \kappa_0 v^{6} v_{\mathrm{h}}^{2}}{3584 q v_{\mathrm{h}}^{2} - 7168 q v_{\mathrm{h}} v_{\mathrm{q}} + 3584 q v_{\mathrm{q}}^{2}}
\\
&\quad - \frac{15 \Delta\Lambda_{\mathrm{HQ}} \kappa_0 v^{6} v_{\mathrm{h}}}{- 16 q v_{\mathrm{h}}^{3} + 48 q v_{\mathrm{h}}^{2} v_{\mathrm{q}} - 48 q v_{\mathrm{h}} v_{\mathrm{q}}^{2} + 16 q v_{\mathrm{q}}^{3}}
- \frac{15 \Delta\Lambda_{\mathrm{HQ}} \kappa_0 v^{6}}{32 q v_{\mathrm{h}}^{2} - 64 q v_{\mathrm{h}} v_{\mathrm{q}} + 32 q v_{\mathrm{q}}^{2}}
\\
&\quad + \frac{15 \Delta\Lambda_{\mathrm{HQ}} \kappa_0 v^{5} v_{\mathrm{h}}^{2}}{- 16 q v_{\mathrm{h}}^{3} + 48 q v_{\mathrm{h}}^{2} v_{\mathrm{q}} - 48 q v_{\mathrm{h}} v_{\mathrm{q}}^{2} + 16 q v_{\mathrm{q}}^{3}}
+ \frac{15 \Delta\Lambda_{\mathrm{HQ}} \kappa_0 v^{5} v_{\mathrm{h}}}{16 q v_{\mathrm{h}}^{2} - 32 q v_{\mathrm{h}} v_{\mathrm{q}} + 16 q v_{\mathrm{q}}^{2}}
\\
&\quad - \frac{5 \Delta\Lambda_{\mathrm{HQ}} \kappa_0 v^{4} v_{\mathrm{h}}^{3}}{- 16 q v_{\mathrm{h}}^{3} + 48 q v_{\mathrm{h}}^{2} v_{\mathrm{q}} - 48 q v_{\mathrm{h}} v_{\mathrm{q}}^{2} + 16 q v_{\mathrm{q}}^{3}}
- \frac{15 \Delta\Lambda_{\mathrm{HQ}} \kappa_0 v^{4} v_{\mathrm{h}}^{2}}{32 q v_{\mathrm{h}}^{2} - 64 q v_{\mathrm{h}} v_{\mathrm{q}} + 32 q v_{\mathrm{q}}^{2}}
\\
&\quad + \hat{a} \Biggl(
- \frac{565 \Delta\Lambda_{\mathrm{HQ}} h_2 \kappa_0 v^{9} v_{\mathrm{h}}^{3}}{- 192 q v_{\mathrm{h}}^{3} + 576 q v_{\mathrm{h}}^{2} v_{\mathrm{q}} - 576 q v_{\mathrm{h}} v_{\mathrm{q}}^{2} + 192 q v_{\mathrm{q}}^{3}}
- \frac{565 \Delta\Lambda_{\mathrm{HQ}} h_2 \kappa_0 v^{9} v_{\mathrm{h}}^{2}}{128 q v_{\mathrm{h}}^{2} - 256 q v_{\mathrm{h}} v_{\mathrm{q}} + 128 q v_{\mathrm{q}}^{2}}
\\
&\qquad\qquad
- \frac{3715 \Delta\Lambda_{\mathrm{HQ}} h_{3s} \kappa_0 v^{9} v_{\mathrm{h}}^{3}}{- 5376 q v_{\mathrm{h}}^{3} + 16128 q v_{\mathrm{h}}^{2} v_{\mathrm{q}} - 16128 q v_{\mathrm{h}} v_{\mathrm{q}}^{2} + 5376 q v_{\mathrm{q}}^{3}}
- \frac{3715 \Delta\Lambda_{\mathrm{HQ}} h_{3s} \kappa_0 v^{9} v_{\mathrm{h}}^{2}}{3584 q v_{\mathrm{h}}^{2} - 7168 q v_{\mathrm{h}} v_{\mathrm{q}} + 3584 q v_{\mathrm{q}}^{2}}
\\
&\qquad\qquad
- \frac{15 \Delta\Lambda_{\mathrm{HQ}} h_{3s} \kappa_0 v^{9} v_{\mathrm{h}}}{- 16 q v_{\mathrm{h}}^{3} + 48 q v_{\mathrm{h}}^{2} v_{\mathrm{q}} - 48 q v_{\mathrm{h}} v_{\mathrm{q}}^{2} + 16 q v_{\mathrm{q}}^{3}}
- \frac{15 \Delta\Lambda_{\mathrm{HQ}} h_{3s} \kappa_0 v^{9}}{32 q v_{\mathrm{h}}^{2} - 64 q v_{\mathrm{h}} v_{\mathrm{q}} + 32 q v_{\mathrm{q}}^{2}}
\\
&\qquad\qquad
+ \frac{15 \Delta\Lambda_{\mathrm{HQ}} h_{3s} \kappa_0 v^{8} v_{\mathrm{h}}^{2}}{- 16 q v_{\mathrm{h}}^{3} + 48 q v_{\mathrm{h}}^{2} v_{\mathrm{q}} - 48 q v_{\mathrm{h}} v_{\mathrm{q}}^{2} + 16 q v_{\mathrm{q}}^{3}}
+ \frac{15 \Delta\Lambda_{\mathrm{HQ}} h_{3s} \kappa_0 v^{8} v_{\mathrm{h}}}{16 q v_{\mathrm{h}}^{2} - 32 q v_{\mathrm{h}} v_{\mathrm{q}} + 16 q v_{\mathrm{q}}^{2}}
\\
&\qquad\qquad
- \frac{5 \Delta\Lambda_{\mathrm{HQ}} h_{3s} \kappa_0 v^{7} v_{\mathrm{h}}^{3}}{- 16 q v_{\mathrm{h}}^{3} + 48 q v_{\mathrm{h}}^{2} v_{\mathrm{q}} - 48 q v_{\mathrm{h}} v_{\mathrm{q}}^{2} + 16 q v_{\mathrm{q}}^{3}}
- \frac{15 \Delta\Lambda_{\mathrm{HQ}} h_{3s} \kappa_0 v^{7} v_{\mathrm{h}}^{2}}{32 q v_{\mathrm{h}}^{2} - 64 q v_{\mathrm{h}} v_{\mathrm{q}} + 32 q v_{\mathrm{q}}^{2}}
\\
&\qquad\qquad
- \frac{732985 \Delta\Lambda_{\mathrm{HQ}} \kappa_0 v^{9} v_{\mathrm{h}}^{3}}{- 32256 q v_{\mathrm{h}}^{3} + 96768 q v_{\mathrm{h}}^{2} v_{\mathrm{q}} - 96768 q v_{\mathrm{h}} v_{\mathrm{q}}^{2} + 32256 q v_{\mathrm{q}}^{3}}
- \frac{732985 \Delta\Lambda_{\mathrm{HQ}} \kappa_0 v^{9} v_{\mathrm{h}}^{2}}{21504 q v_{\mathrm{h}}^{2} - 43008 q v_{\mathrm{h}} v_{\mathrm{q}} + 21504 q v_{\mathrm{q}}^{2}}
\\
&\qquad\qquad
- \frac{565 \Delta\Lambda_{\mathrm{HQ}} \kappa_0 v^{9} v_{\mathrm{h}}}{- 64 q v_{\mathrm{h}}^{3} + 192 q v_{\mathrm{h}}^{2} v_{\mathrm{q}} - 192 q v_{\mathrm{h}} v_{\mathrm{q}}^{2} + 64 q v_{\mathrm{q}}^{3}}
- \frac{565 \Delta\Lambda_{\mathrm{HQ}} \kappa_0 v^{9}}{128 q v_{\mathrm{h}}^{2} - 256 q v_{\mathrm{h}} v_{\mathrm{q}} + 128 q v_{\mathrm{q}}^{2}}
\\
&\qquad\qquad
+ \frac{565 \Delta\Lambda_{\mathrm{HQ}} \kappa_0 v^{8} v_{\mathrm{h}}^{2}}{- 64 q v_{\mathrm{h}}^{3} + 192 q v_{\mathrm{h}}^{2} v_{\mathrm{q}} - 192 q v_{\mathrm{h}} v_{\mathrm{q}}^{2} + 64 q v_{\mathrm{q}}^{3}}
+ \frac{565 \Delta\Lambda_{\mathrm{HQ}} \kappa_0 v^{8} v_{\mathrm{h}}}{64 q v_{\mathrm{h}}^{2} - 128 q v_{\mathrm{h}} v_{\mathrm{q}} + 64 q v_{\mathrm{q}}^{2}}
\\
&\qquad\qquad
- \frac{565 \Delta\Lambda_{\mathrm{HQ}} \kappa_0 v^{7} v_{\mathrm{h}}^{3}}{- 192 q v_{\mathrm{h}}^{3} + 576 q v_{\mathrm{h}}^{2} v_{\mathrm{q}} - 576 q v_{\mathrm{h}} v_{\mathrm{q}}^{2} + 192 q v_{\mathrm{q}}^{3}}
- \frac{565 \Delta\Lambda_{\mathrm{HQ}} \kappa_0 v^{7} v_{\mathrm{h}}^{2}}{128 q v_{\mathrm{h}}^{2} - 256 q v_{\mathrm{h}} v_{\mathrm{q}} + 128 q v_{\mathrm{q}}^{2}}
\Biggr)
\\
&\quad + h_2 \Biggl(
\frac{5 \pi \Delta\Lambda_{\mathrm{HQ}} \kappa_0 v^{9} v_{\mathrm{h}}^{3}}{- 4 q v_{\mathrm{h}}^{3} + 12 q v_{\mathrm{h}}^{2} v_{\mathrm{q}} - 12 q v_{\mathrm{h}} v_{\mathrm{q}}^{2} + 4 q v_{\mathrm{q}}^{3}}
+ \frac{3715 \Delta\Lambda_{\mathrm{HQ}} \kappa_0 v^{9} v_{\mathrm{h}}^{2}}{- 1792 q v_{\mathrm{h}}^{3} + 5376 q v_{\mathrm{h}}^{2} v_{\mathrm{q}} - 5376 q v_{\mathrm{h}} v_{\mathrm{q}}^{2} + 1792 q v_{\mathrm{q}}^{3}}
\\
&\qquad\qquad
+ \frac{15 \pi \Delta\Lambda_{\mathrm{HQ}} \kappa_0 v^{9} v_{\mathrm{h}}^{2}}{8 q v_{\mathrm{h}}^{2} - 16 q v_{\mathrm{h}} v_{\mathrm{q}} + 8 q v_{\mathrm{q}}^{2}}
+ \frac{3715 \Delta\Lambda_{\mathrm{HQ}} \kappa_0 v^{9} v_{\mathrm{h}}}{1792 q v_{\mathrm{h}}^{2} - 3584 q v_{\mathrm{h}} v_{\mathrm{q}} + 1792 q v_{\mathrm{q}}^{2}}
\\
&\qquad\qquad
+ \frac{5 \Delta\Lambda_{\mathrm{HQ}} \kappa_0 v^{9}}{- 16 q v_{\mathrm{h}}^{3} + 48 q v_{\mathrm{h}}^{2} v_{\mathrm{q}} - 48 q v_{\mathrm{h}} v_{\mathrm{q}}^{2} + 16 q v_{\mathrm{q}}^{3}}
- \frac{3715 \Delta\Lambda_{\mathrm{HQ}} \kappa_0 v^{8} v_{\mathrm{h}}^{3}}{- 5376 q v_{\mathrm{h}}^{3} + 16128 q v_{\mathrm{h}}^{2} v_{\mathrm{q}} - 16128 q v_{\mathrm{h}} v_{\mathrm{q}}^{2} + 5376 q v_{\mathrm{q}}^{3}}
\\
&\qquad\qquad
- \frac{3715 \Delta\Lambda_{\mathrm{HQ}} \kappa_0 v^{8} v_{\mathrm{h}}^{2}}{3584 q v_{\mathrm{h}}^{2} - 7168 q v_{\mathrm{h}} v_{\mathrm{q}} + 3584 q v_{\mathrm{q}}^{2}}
- \frac{15 \Delta\Lambda_{\mathrm{HQ}} \kappa_0 v^{8} v_{\mathrm{h}}}{- 16 q v_{\mathrm{h}}^{3} + 48 q v_{\mathrm{h}}^{2} v_{\mathrm{q}} - 48 q v_{\mathrm{h}} v_{\mathrm{q}}^{2} + 16 q v_{\mathrm{q}}^{3}}
\\
&\qquad\qquad
- \frac{15 \Delta\Lambda_{\mathrm{HQ}} \kappa_0 v^{8}}{32 q v_{\mathrm{h}}^{2} - 64 q v_{\mathrm{h}} v_{\mathrm{q}} + 32 q v_{\mathrm{q}}^{2}}
+ \frac{15 \Delta\Lambda_{\mathrm{HQ}} \kappa_0 v^{7} v_{\mathrm{h}}^{2}}{- 16 q v_{\mathrm{h}}^{3} + 48 q v_{\mathrm{h}}^{2} v_{\mathrm{q}} - 48 q v_{\mathrm{h}} v_{\mathrm{q}}^{2} + 16 q v_{\mathrm{q}}^{3}}
\\
&\qquad\qquad
+ \frac{15 \Delta\Lambda_{\mathrm{HQ}} \kappa_0 v^{7} v_{\mathrm{h}}}{16 q v_{\mathrm{h}}^{2} - 32 q v_{\mathrm{h}} v_{\mathrm{q}} + 16 q v_{\mathrm{q}}^{2}}
- \frac{5 \Delta\Lambda_{\mathrm{HQ}} \kappa_0 v^{6} v_{\mathrm{h}}^{3}}{- 16 q v_{\mathrm{h}}^{3} + 48 q v_{\mathrm{h}}^{2} v_{\mathrm{q}} - 48 q v_{\mathrm{h}} v_{\mathrm{q}}^{2} + 16 q v_{\mathrm{q}}^{3}}
\\
&\qquad\qquad
- \frac{15 \Delta\Lambda_{\mathrm{HQ}} \kappa_0 v^{6} v_{\mathrm{h}}^{2}}{32 q v_{\mathrm{h}}^{2} - 64 q v_{\mathrm{h}} v_{\mathrm{q}} + 32 q v_{\mathrm{q}}^{2}}
\Biggr)
\\
&\quad + h_4 \Biggl(
\frac{15 \Delta\Lambda_{\mathrm{HQ}} \kappa_0 v^{9} v_{\mathrm{h}}^{2}}{- 16 q v_{\mathrm{h}}^{3} + 48 q v_{\mathrm{h}}^{2} v_{\mathrm{q}} - 48 q v_{\mathrm{h}} v_{\mathrm{q}}^{2} + 16 q v_{\mathrm{q}}^{3}}
+ \frac{15 \Delta\Lambda_{\mathrm{HQ}} \kappa_0 v^{9} v_{\mathrm{h}}}{16 q v_{\mathrm{h}}^{2} - 32 q v_{\mathrm{h}} v_{\mathrm{q}} + 16 q v_{\mathrm{q}}^{2}}
\\
&\qquad\qquad
- \frac{5 \Delta\Lambda_{\mathrm{HQ}} \kappa_0 v^{8} v_{\mathrm{h}}^{3}}{- 16 q v_{\mathrm{h}}^{3} + 48 q v_{\mathrm{h}}^{2} v_{\mathrm{q}} - 48 q v_{\mathrm{h}} v_{\mathrm{q}}^{2} + 16 q v_{\mathrm{q}}^{3}}
- \frac{15 \Delta\Lambda_{\mathrm{HQ}} \kappa_0 v^{8} v_{\mathrm{h}}^{2}}{32 q v_{\mathrm{h}}^{2} - 64 q v_{\mathrm{h}} v_{\mathrm{q}} + 32 q v_{\mathrm{q}}^{2}}
\Biggr).
\end{aligned}
\end{equation}
\endgroup

Carrying out the integration over the mixed interval and the tail region yields the full truncated closed-form result
\begin{equation}
\scriptsize
\begin{aligned}
\Delta\Phi_{\mathrm{mix}}^{(v^9)}
&= \frac{\Delta\Lambda_{\mathrm{HQ}} \kappa_0}{292626432 q \left(v_{\mathrm{h}}^{3} - 3 v_{\mathrm{h}}^{2} v_{\mathrm{q}} + 3 v_{\mathrm{h}} v_{\mathrm{q}}^{2} - v_{\mathrm{q}}^{3}\right)}
\Biggl(
v_{\mathrm{h}}^{10} \Bigl(
43055712 \hat{a} h_2 v_{\mathrm{h}}^{3}
- 129167136 \hat{a} h_2 v_{\mathrm{h}}^{2} v_{\mathrm{q}}
\\
&\qquad\qquad
+ 10110744 \hat{a} h_{3s} v_{\mathrm{h}}^{3}
- 30332232 \hat{a} h_{3s} v_{\mathrm{h}}^{2} v_{\mathrm{q}}
- 13716864 \hat{a} h_{3s} v_{\mathrm{h}}
- 13716864 \hat{a} h_{3s} v_{\mathrm{q}}
\\
&\qquad\qquad
+ 332481996 \hat{a} v_{\mathrm{h}}^{3}
- 997445988 \hat{a} v_{\mathrm{h}}^{2} v_{\mathrm{q}}
- 129167136 \hat{a} v_{\mathrm{h}}
- 129167136 \hat{a} v_{\mathrm{q}}
\\
&\qquad\qquad
- 18289152 \pi h_2 v_{\mathrm{h}}^{3}
+ 54867456 \pi h_2 v_{\mathrm{h}}^{2} v_{\mathrm{q}}
+ 60664464 h_2 v_{\mathrm{h}} v_{\mathrm{q}}
+ 9144576 h_2
\\
&\qquad\qquad
+ 27433728 h_4 v_{\mathrm{h}} v_{\mathrm{q}}
- 52588116 \pi v_{\mathrm{h}}^{3}
+ 157764348 \pi v_{\mathrm{h}}^{2} v_{\mathrm{q}}
+ 82584171 v_{\mathrm{h}} v_{\mathrm{q}}
\\
&\qquad\qquad
+ 54867456 \pi v_{\mathrm{h}}
+ 54867456 \pi v_{\mathrm{q}}
+ 20221488
\Bigr)
\\
&\quad
+ v_{\mathrm{h}}^{9} \Bigl(
30481920 \hat{a} h_{3s} v_{\mathrm{h}} v_{\mathrm{q}}
+ 287038080 \hat{a} v_{\mathrm{h}} v_{\mathrm{q}}
+ 11234160 h_2 v_{\mathrm{h}}^{3}
\\
&\qquad\qquad
- 33702480 h_2 v_{\mathrm{h}}^{2} v_{\mathrm{q}}
- 15240960 h_2 v_{\mathrm{h}}
- 15240960 h_2 v_{\mathrm{q}}
+ 5080320 h_4 v_{\mathrm{h}}^{3}
\\
&\qquad\qquad
- 15240960 h_4 v_{\mathrm{h}}^{2} v_{\mathrm{q}}
+ 15293365 v_{\mathrm{h}}^{3}
- 45880095 v_{\mathrm{h}}^{2} v_{\mathrm{q}}
- 121927680 \pi v_{\mathrm{h}} v_{\mathrm{q}}
\\
&\qquad\qquad
- 33702480 v_{\mathrm{h}}
- 33702480 v_{\mathrm{q}}
\Bigr)
\\
&\quad
+ v_{\mathrm{h}}^{8} \Bigl(
5715360 \hat{a} h_{3s} v_{\mathrm{h}}^{3}
- 17146080 \hat{a} h_{3s} v_{\mathrm{h}}^{2} v_{\mathrm{q}}
+ 53819640 \hat{a} v_{\mathrm{h}}^{3}
\\
&\qquad\qquad
- 161458920 \hat{a} v_{\mathrm{h}}^{2} v_{\mathrm{q}}
+ 34292160 h_2 v_{\mathrm{h}} v_{\mathrm{q}}
- 22861440 \pi v_{\mathrm{h}}^{3}
+ 68584320 \pi v_{\mathrm{h}}^{2} v_{\mathrm{q}}
\\
&\qquad\qquad
+ 75830580 v_{\mathrm{h}} v_{\mathrm{q}}
+ 11430720
\Bigr)
\\
&\quad
+ v_{\mathrm{h}}^{7} \Bigl(
6531840 h_2 v_{\mathrm{h}}^{3}
- 19595520 h_2 v_{\mathrm{h}}^{2} v_{\mathrm{q}}
+ 14443920 v_{\mathrm{h}}^{3}
- 43331760 v_{\mathrm{h}}^{2} v_{\mathrm{q}}
\\
&\qquad\qquad
- 19595520 v_{\mathrm{h}}
- 19595520 v_{\mathrm{q}}
\Bigr)
\\
&\quad
+ 9144576 v_{\mathrm{h}} \Bigl(
5 v_{\mathrm{h}}^{6} v_{\mathrm{q}}
+ v_{\mathrm{h}}^{6} \left(v_{\mathrm{h}} - 3 v_{\mathrm{q}}\right)
- v_{\mathrm{h}} v_{\mathrm{q}}^{5} \left(v_{\mathrm{h}} - 3 v_{\mathrm{q}}\right)
- 5 v_{\mathrm{q}}^{7}
\Bigr)
\\
&\quad
+ v_{\mathrm{q}}^{10} \Bigl(
- 43055712 \hat{a} h_2 v_{\mathrm{h}}^{3}
+ 129167136 \hat{a} h_2 v_{\mathrm{h}}^{2} v_{\mathrm{q}}
- 10110744 \hat{a} h_{3s} v_{\mathrm{h}}^{3}
\\
&\qquad\qquad
+ 30332232 \hat{a} h_{3s} v_{\mathrm{h}}^{2} v_{\mathrm{q}}
+ 13716864 \hat{a} h_{3s} v_{\mathrm{h}}
+ 13716864 \hat{a} h_{3s} v_{\mathrm{q}}
- 332481996 \hat{a} v_{\mathrm{h}}^{3}
\\
&\qquad\qquad
+ 997445988 \hat{a} v_{\mathrm{h}}^{2} v_{\mathrm{q}}
+ 129167136 \hat{a} v_{\mathrm{h}}
+ 129167136 \hat{a} v_{\mathrm{q}}
+ 18289152 \pi h_2 v_{\mathrm{h}}^{3}
\\
&\qquad\qquad
- 54867456 \pi h_2 v_{\mathrm{h}}^{2} v_{\mathrm{q}}
- 60664464 h_2 v_{\mathrm{h}} v_{\mathrm{q}}
- 9144576 h_2
- 27433728 h_4 v_{\mathrm{h}} v_{\mathrm{q}}
\\
&\qquad\qquad
+ 52588116 \pi v_{\mathrm{h}}^{3}
- 157764348 \pi v_{\mathrm{h}}^{2} v_{\mathrm{q}}
- 82584171 v_{\mathrm{h}} v_{\mathrm{q}}
- 54867456 \pi v_{\mathrm{h}}
\\
&\qquad\qquad
- 54867456 \pi v_{\mathrm{q}}
- 20221488
\Bigr)
\\
&\quad
+ v_{\mathrm{q}}^{9} \Bigl(
- 30481920 \hat{a} h_{3s} v_{\mathrm{h}} v_{\mathrm{q}}
- 287038080 \hat{a} v_{\mathrm{h}} v_{\mathrm{q}}
- 11234160 h_2 v_{\mathrm{h}}^{3}
\\
&\qquad\qquad
+ 33702480 h_2 v_{\mathrm{h}}^{2} v_{\mathrm{q}}
+ 15240960 h_2 v_{\mathrm{h}}
+ 15240960 h_2 v_{\mathrm{q}}
- 5080320 h_4 v_{\mathrm{h}}^{3}
\\
&\qquad\qquad
+ 15240960 h_4 v_{\mathrm{h}}^{2} v_{\mathrm{q}}
- 15293365 v_{\mathrm{h}}^{3}
+ 45880095 v_{\mathrm{h}}^{2} v_{\mathrm{q}}
+ 121927680 \pi v_{\mathrm{h}} v_{\mathrm{q}}
\\
&\qquad\qquad
+ 33702480 v_{\mathrm{h}}
+ 33702480 v_{\mathrm{q}}
\Bigr)
\\
&\quad
+ v_{\mathrm{q}}^{8} \Bigl(
- 5715360 \hat{a} h_{3s} v_{\mathrm{h}}^{3}
+ 17146080 \hat{a} h_{3s} v_{\mathrm{h}}^{2} v_{\mathrm{q}}
- 53819640 \hat{a} v_{\mathrm{h}}^{3}
\\
&\qquad\qquad
+ 161458920 \hat{a} v_{\mathrm{h}}^{2} v_{\mathrm{q}}
- 34292160 h_2 v_{\mathrm{h}} v_{\mathrm{q}}
+ 22861440 \pi v_{\mathrm{h}}^{3}
- 68584320 \pi v_{\mathrm{h}}^{2} v_{\mathrm{q}}
\\
&\qquad\qquad
- 75830580 v_{\mathrm{h}} v_{\mathrm{q}}
- 11430720
\Bigr)
\\
&\quad
+ v_{\mathrm{q}}^{7} \Bigl(
- 6531840 h_2 v_{\mathrm{h}}^{3}
+ 19595520 h_2 v_{\mathrm{h}}^{2} v_{\mathrm{q}}
- 14443920 v_{\mathrm{h}}^{3}
+ 43331760 v_{\mathrm{h}}^{2} v_{\mathrm{q}}
\\
&\qquad\qquad
+ 19595520 v_{\mathrm{h}}
+ 19595520 v_{\mathrm{q}}
\Bigr)
\\
&\quad
- \left(v_{\mathrm{h}}^{3} - 3 v_{\mathrm{h}}^{2} v_{\mathrm{q}} + 3 v_{\mathrm{h}} v_{\mathrm{q}}^{2} - v_{\mathrm{q}}^{3}\right)
\Bigl(
v_{\mathrm{ISCO}}^{10} \Bigl(
43055712 \hat{a} h_2
+ 10110744 \hat{a} h_{3s}
\\
&\qquad\qquad
+ 332481996 \hat{a}
- 18289152 \pi h_2
- 52588116 \pi
\Bigr)
+ v_{\mathrm{ISCO}}^{9} \Bigl(
11234160 h_2
+ 5080320 h_4
+ 15293365
\Bigr)
\\
&\qquad\qquad
+ v_{\mathrm{ISCO}}^{8} \Bigl(
5715360 \hat{a} h_{3s}
+ 53819640 \hat{a}
- 22861440 \pi
\Bigr)
+ v_{\mathrm{ISCO}}^{7} \Bigl(
6531840 h_2
+ 14443920
\Bigr)
\\
&\qquad\qquad
+ 9144576 v_{\mathrm{ISCO}}^{5}
+ v_{\mathrm{q}}^{10} \Bigl(
- 43055712 \hat{a} h_2
- 10110744 \hat{a} h_{3s}
- 332481996 \hat{a}
\\
&\qquad\qquad\qquad
+ 18289152 \pi h_2
+ 52588116 \pi
\Bigr)
+ v_{\mathrm{q}}^{9} \Bigl(
- 11234160 h_2
- 5080320 h_4
- 15293365
\Bigr)
\\
&\qquad\qquad
+ v_{\mathrm{q}}^{8} \Bigl(
- 5715360 \hat{a} h_{3s}
- 53819640 \hat{a}
+ 22861440 \pi
\Bigr)
+ v_{\mathrm{q}}^{7} \Bigl(
- 6531840 h_2
- 14443920
\Bigr)
\\
&\qquad\qquad
- 9144576 v_{\mathrm{q}}^{5}
\Bigr)
\Biggr).
\end{aligned}
\end{equation}

At leading order, the mixed-phase result simplifies to
\begin{equation}
\Delta\Phi_{\rm mix}^{\rm LO}
=
\frac{\Delta\Lambda_{HQ}\kappa_{0}}{896q}
\left(
-28v_{\rm ISCO}^{5}
+3v_h^{5}
+5v_h^{4}v_q
+6v_h^{3}v_q^{2}
+6v_h^{2}v_q^{3}
+5v_hv_q^{4}
+3v_q^{5}
\right).
\label{eq:app_DPhi_mix_LO}
\end{equation}
This expression makes explicit how a finite transition width smooths the sharp-jump result into a polynomial average over the interval endpoints.

\subsection{Consistency of the two limits}

The sharp-transition and mixed-phase constructions are continuously connected. In the limit
\begin{equation}
v_h \rightarrow v_q \rightarrow v_c,
\end{equation}
the mixed-phase profile collapses to a Heaviside jump, and the mixed-phase leading-order expression reduces exactly to the sharp-transition result. The two dephasing models should therefore be understood not as disconnected alternatives, but as two limiting realizations of the same effective tidal-response framework.


\section*{References}

\begin{thebibliography}{99}

\bibitem{LISAESA}
European Space Agency, ``LISA -- Laser Interferometer Space Antenna,''
available at \url{https://www.cosmos.esa.int/web/lisa}
(accessed 2026-03-12).

\bibitem{BardeenPressTeukolsky1972}
J. M. Bardeen, W. H. Press, and S. A. Teukolsky,
``Rotating Black Holes: Locally Nonrotating Frames, Energy Extraction, and Scalar Synchrotron Radiation,''
\textit{Astrophys. J.} \textbf{178}, 347--370 (1972).
doi:10.1086/151796.

\bibitem{Teukolsky1973}
S. A. Teukolsky,
``Perturbations of a Rotating Black Hole. I. Fundamental Equations for Gravitational, Electromagnetic, and Neutrino-Field Perturbations,''
\textit{Astrophys. J.} \textbf{185}, 635--647 (1973).
doi:10.1086/152444.

\bibitem{Hughes2000}
S. A. Hughes,
``The Evolution of Circular, Nonequatorial Orbits of Kerr Black Holes due to Gravitational-Wave Emission,''
\textit{Phys. Rev. D} \textbf{61}, 084004 (2000).
doi:10.1103/PhysRevD.61.084004.

\bibitem{Hughes2001}
S. A. Hughes,
``Evolution of Circular, Nonequatorial Orbits of Kerr Black Holes due to Gravitational-Wave Emission. II. Inspiral Trajectories and Gravitational Waveforms,''
\textit{Phys. Rev. D} \textbf{64}, 064004 (2001).
doi:10.1103/PhysRevD.64.064004.

\bibitem{BarackPound2019}
L. Barack and A. Pound,
``Self-force and Radiation Reaction in General Relativity,''
\textit{Rep. Prog. Phys.} \textbf{82}, 016904 (2019).
doi:10.1088/1361-6633/aae552.

\bibitem{Fujita2015}
R. Fujita,
``Gravitational Waves from a Particle in Circular Orbits around a Rotating Black Hole to the 11th Post-Newtonian Order,''
\textit{Prog. Theor. Exp. Phys.} \textbf{2015}, 033E01 (2015).
doi:10.1093/ptep/ptv024.

\bibitem{Blanchet2024PN}
L. Blanchet,
``Gravitational Radiation from Post-Newtonian Sources and Inspiralling Compact Binaries,''
\textit{Living Rev. Relativ.} \textbf{17}, 2 (2014).
doi:10.12942/lrr-2014-2.

\bibitem{FengLyuYang2021}
X. Feng, Z. Lyu, and H. Yang,
``Black-Hole Perturbation Plus Post-Newtonian Theory: Hybrid Waveform for Neutron Star Binaries,''
arXiv:2104.11848 [gr-qc] (2021).

\bibitem{FengLyuYang2022}
X. Feng, Z. Lyu, and H. Yang,
``Black-Hole Perturbation Plus Post-Newtonian Theory: Hybrid Waveform for Neutron Star Binaries,''
\textit{Phys. Rev. D} \textbf{105}, 104043 (2022).
doi:10.1103/PhysRevD.105.104043.

\bibitem{Dolan2024}
S. R. Dolan, L. Durkan, C. Kavanagh, and B. Wardell,
``Metric Perturbations of Kerr Spacetime in Lorenz Gauge: Circular Equatorial Orbits,''
\textit{Class. Quantum Grav.} \textbf{41}, 155011 (2024).
doi:10.1088/1361-6382/ad52e3.

\bibitem{Khalvati2024}
H. Khalvati, A. Santini, F. Duque, L. Speri, J. Gair, H. Yang, and R. Brito,
``Impact of Relativistic Waveforms in LISA's Science Objectives with Extreme-Mass-Ratio Inspirals,''
\textit{Phys. Rev. D} \textbf{111}, 082010 (2025).
doi:10.1103/PhysRevD.111.082010.

\bibitem{SearchEMRI2025}
S. H. Strub, L. Speri, and D. Giardini,
``Searching for Extreme Mass Ratio Inspirals in LISA: From Identification to Parameter Estimation,''
arXiv:2505.17814 [gr-qc] (2025).

\bibitem{Badger2024}
C. Badger, J. A. Font, M. Sakellariadou, and A. Torres-Forn{\'e},
``High-Speed Reconstruction of Long-Duration Gravitational Waves from Extreme Mass Ratio Inspirals Using Sparse Dictionary Learning,''
\textit{Phys. Rev. D} \textbf{110}, 064074 (2024).
doi:10.1103/PhysRevD.110.064074.

\bibitem{HanSteiner2019}
S. Han and A. W. Steiner,
``Tidal Deformability with Sharp Phase Transitions in (Binary) Neutron Stars,''
\textit{Phys. Rev. D} \textbf{99}, 083014 (2019).
doi:10.1103/PhysRevD.99.083014.

\bibitem{Takatsy2023}
J. Tak{\'a}tsy, P. Kov{\'a}cs, G. Wolf, and J. Schaffner-Bielich,
``What Neutron Stars Tell about the Hadron--Quark Phase Transition: A Bayesian Study,''
\textit{Phys. Rev. D} \textbf{108}, 043002 (2023).
doi:10.1103/PhysRevD.108.043002.

\bibitem{RaithelMost2023PRL}
C. A. Raithel and E. R. Most,
``Degeneracy in the Inference of Phase Transitions in the Neutron Star Equation of State from Gravitational Wave Data,''
\textit{Phys. Rev. Lett.} \textbf{130}, 201403 (2023).
doi:10.1103/PhysRevLett.130.201403.

\bibitem{RaithelMost2023PRD}
C. A. Raithel and E. R. Most,
``Tidal Deformability Doppelgangers: Implications of a Low-Density Phase Transition in the Neutron Star Equation of State,''
\textit{Phys. Rev. D} \textbf{108}, 023010 (2023).
doi:10.1103/PhysRevD.108.023010.

\bibitem{Roy2024}
D. G. Roy, A. Venneti, T. Malik, S. Bhattacharya, and S. Banik,
``Bayesian Evaluation of Hadron--Quark Phase Transition Models through Neutron Star Observables in Light of Nuclear and Astrophysics Data,''
\textit{Phys. Lett. B} \textbf{859}, 139128 (2024).
doi:10.1016/j.physletb.2024.139128.

\bibitem{Li2025QCDPT}
R. Li, S. Han, Z. Lin, L. Wang, K. Zhou, and S. Shi,
``Toward Constraining QCD Phase Transitions in Neutron Star Interiors: Bayesian Inference with a Tolman--Oppenheimer--Volkoff Linear Response Analysis,''
\textit{Phys. Rev. D} \textbf{111}, 074026 (2025).
doi:10.1103/PhysRevD.111.074026.

\bibitem{Counsell2025}
A. R. Counsell, F. Gittins, N. Andersson, and I. Tews,
``Interface Modes in Inspiralling Neutron Stars: A Gravitational-Wave Probe of First-Order Phase Transitions,''
arXiv:2504.06181 [gr-qc] (2025).

\bibitem{Pereira2025}
J. P. Pereira, L. Tonetto, M. Bejger, J. L. Zdunik, and P. Haensel,
``Dynamical Tides in Neutron Stars with First-Order Phase Transitions: The Role of the Discontinuity Mode,''
arXiv:2504.16911 [astro-ph.HE] (2025).

\bibitem{MathewsPound2025}
J. Mathews and A. Pound,
``Post-adiabatic Waveform-Generation Framework for Asymmetric Precessing Binaries,''
arXiv:2501.01413 [gr-qc] (2025).

\end{thebibliography}
\end{document}